\documentclass[%
reprint, 
superscriptaddress, 
amsmath,amssymb,
]{revtex4-2}

\usepackage{xurl}

\usepackage{graphicx}
\usepackage{epstopdf}
\usepackage{subcaption}
\usepackage{dcolumn}
\usepackage{bm}
\usepackage[colorlinks, linkcolor=red, anchorcolor=blue, citecolor=green]{hyperref}
\usepackage{dcolumn}
\usepackage{soul, color, xcolor} 
\usepackage{amsmath}%
\usepackage{float}
\usepackage{pifont} 

\usepackage{braket}
\usepackage{array}
\usepackage{booktabs}             

\usepackage{threeparttable}       
\usepackage{multirow}     
\usepackage{diagbox}    
\usepackage{array}      
\usepackage{caption}    
\usepackage{ragged2e}
\usepackage{ulem}
\usepackage{xspace}
\usepackage{times}

\usepackage{lineno}

\newcommand{\SQRTX}{$X_{\pi/2}$\xspace}

\newcommand{\addressa}{\affiliation{Laboratory of Quantum Information, University of Science and Technology of China, Hefei, Anhui, 230026, China}}
\newcommand{\addressb}{\affiliation{CAS Center For Excellence in Quantum Information and Quantum Physics, University of Science and Technology of China, Hefei, Anhui, 230026, China}}
\newcommand{\addressc}{\affiliation{Institute of Artificial Intelligence, Hefei Comprehensive National Science Center, Hefei, Anhui, 230088, China}}
\newcommand{\addressd}{\affiliation{Institute of the Advanced Technology, University of Science and Technology of China, Hefei, Anhui, 230088, China}}
\newcommand{\addresse}{\affiliation{Origin Quantum, Hefei, Anhui 230088, China}}
\newcommand{\addressf}{\affiliation{Suzhou Institute for Advanced Research, University of Science and Technology of China, Suzhou, Jiangsu 215123, China}}

\begin{document}
	
\preprint{APS/123-QED}
\title{
    Demonstrating Coherent Quantum Routers for Bucket-Brigade Quantum Random Access Memory on a Superconducting Processor
}

\author{Sheng Zhang}
\addressa\addressb\addressf
\author{Yun-Jie Wang}
\addressd
\author{Peng Wang}
\addressa\addressb\addressf
\author{Ren-Ze Zhao}
\author{Xiao-Yan Yang}
\author{Ze-An Zhao}
\author{Tian-Le Wang}
\author{Hai-Feng Zhang}
\author{Zhi-Fei Li}
\author{Yuan Wu}
\addressa\addressb

\author{Hao-Ran Tao}

\author{Liang-Liang Guo}
\author{Lei Du}
\author{Chi Zhang}
\author{Zhi-Long Jia}
\author{Wei-Cheng Kong}
\addresse

\author{Zhuo-Zhi Zhang}
\author{Xiang-Xiang Song}
\addressa\addressb\addressf

\author{Yu-Chun Wu}
\addressa\addressb\addressc

\author{Zhao-Yun Chen}
\email{chenzhaoyun@iai.ustc.edu.cn}
\addressc

\author{Peng Duan}
\email{pengduan@ustc.edu.cn}
\addressa\addressb

\author{Guo-Ping Guo}
\email{gpguo@ustc.edu.cn}
\addressa\addressb\addresse


\date{\today}

\begin{abstract}
    
Quantum routers (QRouters) are essential components of bucket-brigade quantum random access memory (QRAM), enabling quantum applications such as Grover’s search and quantum machine learning. 
Despite significant theoretical advances, achieving scalable and coherent QRouters experimentally remains challenging. 
Here, we demonstrate coherent quantum routers using a superconducting quantum processor, laying a practical foundation for scalable QRAM systems. 
The quantum router at the core of our implementation utilizes the transition composite gate (TCG) scheme, wherein auxiliary energy levels temporarily mediate conditional interactions, substantially reducing circuit depth compared to traditional gate decompositions. Moreover, by encoding routing addresses in the non-adjacent qutrit states $|0\rangle$ and  $|2\rangle$, our design inherently enables eraser-detection capability, providing efficient post-selection to mitigate routing errors. 
Experimentally, we achieve individual QRouter fidelities up to 95.74\%, and validate scalability through a two-layer quantum routing network achieving an average fidelity of 82.40\%. Our results represent a significant advancement in quantum routing technology, providing enhanced fidelity, built-in error resilience, and practical scalability crucial for the development of future QRAM and large-scale quantum computing architectures.

\end{abstract}

\maketitle
	
\section{\label{sec:level1}Introduction}

Quantum random access memory (QRAM) stands as a cornerstone of quantum computing~\cite{giovannetti2008quantum,hann2019hardware,paler2020parallelizing,asaka2021quantum,chen2021scalable,jaques2023qram,weiss2024qram}, enabling superposition-based access to classical data—a crucial capability for algorithms such as Grover’s search~\cite{Grover1997}, quantum machine learning~\cite{biamonte2017quantum,cerezo2022challenges}, and quantum simulation~\cite{Georgescu2014quantum,Kandala2017hardware}.

Among various QRAM designs, the bucket-brigade architecture~\cite{giovannetti2008architectures,giovannetti2008quantum} stands out for its favorable scaling and noise resilience~\cite{hann2021resilience},
which organizes QRAM access as a conditional routing process along a binary-tree network.
The key enabling element in this scheme is the quantum router (QRouter)—a conditional switch that coherently routes a quantum input to one of two output paths based on the superposed state. 
Through recursive composition of such routers, a complete 
binary-tree routing network can be constructed to support superposition-based queries for QRAM.

While the concept of quantum routers has been widely explored in both theory and experiment~\cite{zhou2009quantum,gonzalez2016nonreciprocal,cheng2016single,zhou2013quantum,zhu2019single,lu2015t,lu2014single,wang2021experimental,ren2022nonreciprocal,li2024quantum,du2020nonreciprocal,agarwal2012optomechanical,yuan2015experimental,bartkiewicz2018implementation,palaiodimopoulos2024chiral}, 
most of these studies focus on quantum communication scenarios, where routing typically refers to controlling the 
propagation of flying qubits.
While preserving coherence in transit, they are predominantly routed via nondeterministic~\cite{yuan2015experimental,bartkiewicz2018implementation}, incomplete-transmission methods~\cite{wang2021experimental,li2024quantum}, rendering them unsuitable for the reversible bucket-brigade QRAM.
In contrast, stationary qubit systems, such as superconducting circuits, are particularly well suited to support fully coherent, deterministic, and cascadable routing~\cite{Yan2018Tunable,arute2019quantum,gao2024establishingnewbenchmarkquantum}. 
However, they also face challenges: native multi-qubit routing requires complex initialization and calibration, while digital decomposition deepens circuits, exacerbating decoherence and cumulative gate errors.
These challenges
highlight the need for new architectures that bridge the gap between router-level coherence and QRAM-level scalability.


In this work, we address these challenge by implementing QRouters
based on the recently proposed transition composite gate (TCG) scheme~\cite{zhang2024realization}.
The TCG scheme uses auxiliary energy levels as intermediate states to mediate conditional interactions, allowing for shallower gate sequences than conventional Clifford decompositions. 
This results in improved coherence and reduced circuit overhead—an essential feature for routing networks. 
Furthermore, by encoding the address qubit in the non-adjacent qutrit states \(|0\rangle\) and \(|2\rangle\), we enable an eraser-detection mechanism~\cite{Kubica2023Erasure,Vittal2023ERASER}, 
through which 
enormous routing errors
can be identified and discarded by post-selection.
This provides lightweight error mitigation
without the need for ancillary hardware. 

Leveraging both shallow circuit depth and eraser-detection-based error suppression,
we present the first experimental demonstration of coherent quantum routers for bucket-brigade QRAM on a superconducting quantum processor. Specifically, we experimentally construct and characterize both a single QRouter and a two-layer quantum routing network, with fidelities up to 95.74\% and 82.40\%, respectively. 
These results establish a scalable and error-resilient routing architecture with minimal overhead, representing a significant step toward practical QRAM systems.

\section{QRouter Architecture and Implementation}
    \begin{figure}
    \centering  \includegraphics[width=1.\linewidth]{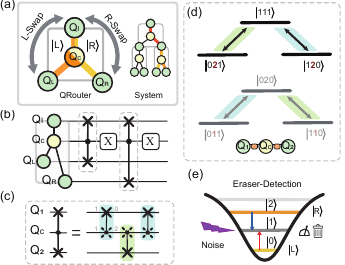}
     \caption{\justifying
        \textbf{Design and operation of the quantum router (QRouter) and routing network.}
        (a) Logical structure of a QRouter and its recursive integration into a binary-tree routing network. Each QRouter routes an input qubit (\(\mathrm{Q_I}\)) to a path qubit (\(\mathrm{Q_L}\) or \(\mathrm{Q_R}\)) based on the address state of \(\mathrm{Q_C}\).  
        (b) Circuit decomposition of the QRouter using two CSWAP gates and two address-flipping operations.  
        (c) Gate-level realization of CSWAP via the TCG scheme, using three \(\sqrt{\text{CZ}}\) gates that induce transitions between \(|11\rangle\) and \(|02\rangle\).  
        (d) Transition pathways for the TCG-based CSWAP under address states \(|1\rangle\) (gray) and \(|2\rangle\) (black).  
        (e) Illustration of routing errors caused by leakage to \(|1\rangle\) and the eraser-detection protocol for error suppression through post-selection.  
        }
		\label{fig 1}
	\end{figure}

A QRouter implements a conditional swap operation between an input qubit \(\mathrm{Q_I}\) and one of two path qubits (\(\mathrm{Q_L}\), \(\mathrm{Q_R}\)), depending on the quantum state of an address qutrit \(\mathrm{Q_C}\). The logical function is defined as
\begin{equation}    
     U_{\mathrm{QRouter}} = |L\rangle \langle L| \otimes \mathrm{Swap}_{\mathrm{Q}_\mathrm{L},\mathrm{Q_I}} + |R\rangle \langle R| \otimes \mathrm{Swap}_{\mathrm{Q}_\mathrm{R},\mathrm{Q}_\mathrm{I}},
\end{equation}
where \(|L\rangle\) and \(|R\rangle\) are two orthogonal address states (e.g., \(|0\rangle\), \(|1\rangle\) or \(|2\rangle\)) encoded in \(\mathrm{Q_C}\). The QRouter thus conditionally transfers the quantum state of \(\mathrm{Q_I}\) to either \(\mathrm{Q_L}\) or \(\mathrm{Q_R}\), realizing address-dependent data routing.

Figure~\ref{fig 1}(a) illustrates the logical structure of a QRouter and how multiple such routers are recursively composed into a binary-tree network to support superposition-based queries. The physical implementation of the QRouter, shown in Fig.~\ref{fig 1}(b), consists of two controlled-SWAP (CSWAP) gates and two address-flipping operations. Each CSWAP conditionally exchanges \(\mathrm{Q_I}\) with \(\mathrm{Q_L}\) or \(\mathrm{Q_R}\), based on the state of \(\mathrm{Q_C}\).

Direct realization of three-qubit CSWAP gates is experimentally demanding. To overcome this, we employ the transition composite gate (TCG) scheme~\cite{zhang2024realization}, which decomposes the CSWAP into a sequence of three two-qubit entangling gates (\(\sqrt{\text{CZ}}\)), as shown in Fig.~\ref{fig 1}(c). The key idea is to temporarily access the non-computational \(|2\rangle\) level of transmons to enable state transitions such as \(|11\rangle \leftrightarrow |20\rangle\) or \(|02\rangle\)~\cite{Foxen2020,Sung2021}, thus implementing effective conditional operations with reduced gate depth.

\begin{table}[tb]
\caption{\justifying
\textbf{Gate counts and circuit depth for different QRouter implementations.}
Comparison between standard Clifford decomposition and the TCG-based QRouter under two encoding schemes: non-eraser (address states \(|0\rangle\), \(|1\rangle\)) and eraser scheme (address states \(|0\rangle\), \(|2\rangle\)). 
TCG significantly reduces the number of single- and two-qubit gates, as well as circuit depth, compared to the Clifford-based implementation. The eraser scheme incurs slightly more single-qubit gates due to address-state flipping, but retains overall efficiency.
}
    \centering
    \begin{tabular}{cccc}
        \toprule[1pt]
        \midrule
        \textbf{QRouter} & \textbf{Clifford} & \textbf{TCG}$^\text{non-eraser}$ & \textbf{TCG}$^\text{eraser}$ \\ \midrule
        \textbf{$N_{\rm 1q}$} & 20 & 2 & 6 \\ 
        \textbf{$N_{\rm 2q}$} & 16 & 6 & 6\\ 	
        \midrule 
        \textbf{Depth} & 30 & 8 & 12\\	
        \midrule
        \bottomrule[1pt]
    \end{tabular}
    \label{tab 1}
\end{table}

The transition pathways for the TCG-based CSWAP are shown in Fig.~\ref{fig 1}(d). A conditional swap can be achieved through sequences like \(|011\rangle \rightarrow |020\rangle \rightarrow |110\rangle\) or \(|021\rangle \leftrightarrow |111\rangle \leftrightarrow |120\rangle\), depending on whether the control qutrit \(\mathrm{Q_C}\) is in \(|1\rangle\) or \(|2\rangle\) (gray and black pathways, respectively). When \(\mathrm{Q_C}\) is in the \(|0\rangle\) state, the swap operation is inhibited.

This TCG-based approach represents a key advantage of our design: by engineering intermediate-level transitions, we achieve a shallower gate decomposition compared to traditional Clifford-based methods, reducing both circuit depth and accumulated error. Table~\ref{tab 1} quantitatively compares gate counts and circuit depths for different decomposition strategies, with further details available in the supplementary material~\cite{supp_QRouter}.

The address qutrit \(\mathrm{Q_C}\) can be encoded in two configurations: \(|L\rangle = |0\rangle\), \(|R\rangle = |1\rangle\) (non-eraser scheme), or \(|L\rangle = |0\rangle\), \(|R\rangle = |2\rangle\) (eraser scheme). The latter enables an additional mechanism for error mitigation where the \(\mathrm{Q_C}\) functions as an eraser qubit. As shown in Fig.~\ref{fig 1}(e), gate imperfections, decoherence, or thermal population can cause \(\mathrm{Q_C}\) to transition into the intermediate \(|1\rangle\) state, which leads to routing errors.
In the eraser scheme, these erroneous events can be post-selected by measuring the final state of \(\mathrm{Q_C}\). If \(\mathrm{Q_C}\) is found in \(|1\rangle\), the corresponding data is discarded. This built-in eraser-detection mechanism allows us to suppress a dominant class of routing errors without introducing additional hardware or qubits—another core advantage of our QRouter design.


We implemented the QRouter using 10 transmon qubits on the superconducting processor \textit{Wukong}, with additional details provided in the supplementary materials~\cite{supp_QRouter}.  
 The \(\sqrt{\text{CZ}}\) gates are realized by tuning frequencies of qubits and coupler close to the $|02\rangle\leftrightarrow|11\rangle$ interaction point wherein these two states accomplish a swap operation.
 The gates are further optimized through Floquet sequence experiments~\cite{arute2020observation,neill2021accurately,gross2024characterizing} (See supplementary materials~\cite{supp_QRouter}). Address preparation and flipping (X) were carried out using single-qubit gates designed with the derivative reduction by adiabatic gate scheme~\cite{motzoi2009simple}. The state $|0\rangle \leftrightarrow |1\rangle$ flip uses an \(X^{01}_{\pi}\) gate, while the state $|1\rangle \leftrightarrow |2\rangle$ flip uses an \(X^{12}_{\pi}\) gate.

All address qubits were measured at the end of each experiment. Unless otherwise specified, we used the eraser scheme encoding.

    \section{\label{sec:level31}Characterization of Quantum router}
	
    \begin{figure}
    \centering
    \includegraphics[width=1.0\linewidth]{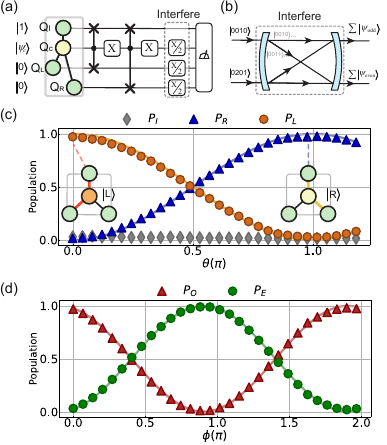}
    \caption{\justifying
        \textbf{Verification of conditional and coherent routing properties of the QRouter.}  
        (a) Experimental setup for testing both transmission (via $\theta$ variation) and coherence (via $\phi$ variation) of the Qrouter with address qubit $\mathrm{Q_C}$ initialized in the state $|\psi_c\rangle = \cos\theta|0\rangle + e^{i\phi}\sin\theta|2\rangle$.  
        (b) Interference scheme using three $X_{\pi/2}$ gates to probe phase-sensitive entanglement.  
        (c) Routing population versus $\theta$, showing expected $\sin^2(\theta)$ and $\cos^2(\theta)$ behavior for path qubits $\mathrm{Q_L}$ and $\mathrm{Q_R}$.  
        (d) Phase-coherent interference oscillations in the population of odd- and even-excitation states as a function of $\phi$, indicating address-path coherence. Dots represent experimental data (5000 shots); solid lines are theoretical predictions.  
        }

    \label{fig 2}
    \end{figure}
    
We begin by characterizing the conditional routing and the coherence properties of the Qrouter, where Fig.~\ref{fig 2} illustrates the experimental setups.
First, we prepared the address qutrit at a superposition state $|\psi_c\rangle=\cos{\theta}|0\rangle_c + e^{i\phi}\sin{\theta}|2\rangle_c$, and observed how information flows from input qubit $\mathrm{Q_I}$ to path qubits $\mathrm{Q_L}$ and $\mathrm{Q_R}$.
As shown in Fig.~\ref{fig 2}(c), we varied \(\theta\) and find that the population distributions of \(P_L\)(\(P_R\)) varied continuously with adjusted \(\theta\), following theoretical  \(\sin^2(\theta)\) (\(\cos^2(\theta)\)) behavior (lines).
The residual population of input qubit (\(P_I\)) stayed below a certain threshold ($2.05\%$) after routing, indicating high transmission efficiency of approximately $98\%$.
Second, to verify the phase coherence during routing, we scanned \(\phi\) and parallelly applied three \SQRTX operations to interfere with the generated entangled states of \(\mathrm{Q}_{\rm I}\), \(\mathrm{Q}_{\rm C}\), \(\mathrm{Q}_{\rm L}\) and \(\mathrm{Q}_{\rm R}\), as illustrated in Fig.~\ref{fig 2}(b). 
The phase of the address changes the phase difference between the entangled states generated under this coherent control~\cite{supp_QRouter}.
This leads to oscillations in the population of odd-excitation states (\(P_{O}=\sum_k P_{|k\rangle}\), \(k\in\{0001,0010,0200,0211\}\)) as \(\sin^2(\phi)\). Even-excitation states (\(P_{E}=\sum_k P_{|k\rangle}\), \(k\in\{0000,0011,0201,0210\}\)) exhibited complementary behaviors, indicating a \(\phi\)-driven coherence process, as shown in Fig.~\ref{fig 2}(d). 
The high-contrast routing in Fig.~\ref{fig 2}(c) and clear interference patterns in Fig.~\ref{fig 2}(d) demonstrate that our QRouter effectively achieves deterministic routing with address-path entanglement.

    \begin{figure}
	\centering
		\includegraphics[width=1.0\linewidth]{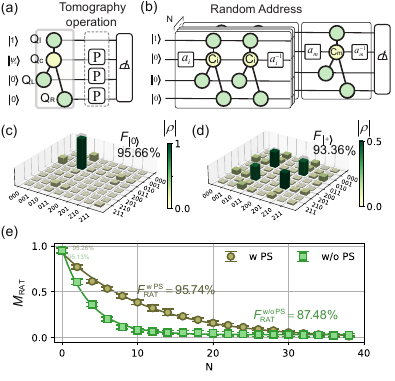}
  \caption{\justifying
        \textbf{Quantum state tomography (QST) and random access test (RAT) of the QRouter.}  
        (a) Experimental circuit used for QST under different addresses. The tomography includes single-qubit basis rotations (P) and simultaneous readout.  
        (b) Circuit structure of the random access test (RAT), consisting of \(N+1\) blocks, with the input qubit initially prepared in the \(|1\rangle\) state. The first \(N\) blocks have random addresses \(a_i\), with two QRouters sharing the same address to implement round-trip information transfer. The final block contains a single QRouter with a random address, transferring the information to data qubits.
        (c,d) Reconstructed density matrices $\rho$ from QST under addresses $|0\rangle$ and $|+\rangle = (|0\rangle + |2\rangle)/\sqrt{2}$, with fidelities $95.66\%$ and $93.36\%$, respectively.  
        (e) Extracted $M_{\mathrm{RAT}}$ values as a function of depth $N$ for eraser and non-eraser schemes. Each point is averaged over 30 experiments with 5000 shots per trial.  
        }

		\label{fig 3}
	\end{figure}
	
    To benchmark the fidelity of our QRouter, we employed a standard quantum state tomography (QST), with the setup shown in Fig.~\ref{fig 3}(a). After routing under various addresses, QST reconstructed the experimental state density matrix ($\rho$) of \(\mathrm{Q}_{\rm{C}}\), \(\mathrm{Q}_{\rm L}\) and \(\mathrm{Q}_{\rm R}\),
    with fidelities as $F_{\mathrm{QST}}=\langle\Psi|\rho|\Psi\rangle$, where $|\Psi\rangle$ is the target state. Figures~\ref{fig 3}(c) and (d) display QST results under addresses $|0\rangle$ and $|+\rangle= (|0\rangle + |2\rangle)/\sqrt{2}$ respectively,
    where corresponding fidelities are $95.66\%$ and $93.36\%$.
    
    We further developed a random access test (RAT) to evaluate the system's average fidelity (see Fig.~\ref{fig 3}(b)).
    Here, random address sequences $\{(a_1,a_1),...,(a_n,a_n),(a_{n+1})\}$ ($a_i\in\{|0\rangle,|2\rangle,|+\rangle,|-\rangle\}$) where $|\pm\rangle = (|0\rangle\pm |2\rangle)/\sqrt{2} $) were loaded into $2N+1$ QRouters, with each sequence repeated $30$ times. 
    Then we measured $\mathrm{Q}_{\rm I}$, $\mathrm{Q}_{\rm L}$ and $\mathrm{Q}_{\rm R}$ to determine the population $P_{|k'\rangle}$ ($k'\in\{000,001,010,011\}$). 
    We regarded the consistency between ideal ($P_{|t\rangle}^0$) and experimental ($P_{|t\rangle}^e$) populations,
    $M_{\rm RAT}=1-{\sum_t |P_{|t\rangle}^0-P_{|t\rangle}^e|}$,
    as the quantitative measure, where $|t\rangle$ is the routing target state.
    We further use $M_{\rm RAT}(N)=l_1+l_2(1-F_{\rm RAT})^{2N+1}$ fitting to extract the fidelity $F_{\rm RAT}$, where $l_1$ and $l_2$ are fitting parameters.
    Ultimately, we obtain a fidelity of 95.74\% under the eraser scheme ($F^{\rm w~PS}_{\rm RAT}$, brown), compared to 87.48\% under the non-eraser scheme ($F^{\rm w/o~PS}_{\rm RAT}$, green), as shown in Fig.~\ref{fig 3}(e).
    
    
This high-fidelity QRouter, endowed with independent routing capabilities, shows strong potential for constructing scalable routing networks.

\section{\label{sec:level32}Two-layer quantum routing network}

\begin{figure*}[ht]
\centering
\includegraphics[width=0.95\linewidth]{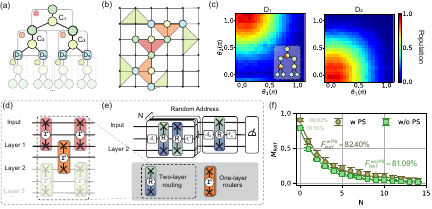}
\caption{\justifying
\textbf{Design and characterization of a two-layer quantum routing network.}
(a) Logical diagram of the binary-tree routing structure used in bucket-brigade QRAM.  
(b) Qubit layout on the superconducting processor, based on a triangular tiling scheme for optimized planar connectivity.  
(c) Measured populations of data qubits \(\rm D_1\) and \(\rm D_2\) as functions of address parameters \((\theta_1, \theta_2)\) encoded in address qutrits \(\rm C_1\) and \(\rm C_2\), demonstrating controlled excitation transfer.  
(d) Simplified circuit structure enabling symmetric top-down and bottom-up routing, supporting layer-wise parallel activation of QRouters.  
(e) Random access test (RAT) protocol for the two-layer network using randomized address sequences across blocks.  
(f) Extracted fidelities from RAT measurements as a function of routing depth \(N\), for both eraser and non-eraser schemes. Each data point is averaged over 30 runs with 5000 shots per trial.
}
\label{fig 4}
\end{figure*}

Cascaded quantum routers form the foundational architecture of bucket-brigade QRAM, where conditional data routing is achieved through a binary-tree network of QRouter units (see Fig.~\ref{fig 4}(a)). While we have demonstrated the performance of individual QRouters, scaling the system to multi-layer networks introduces additional practical challenges. In particular, query efficiency (i.e., the circuit depth and timing required to resolve a path) and layout efficiency (the physical feasibility of embedding the network onto a quantum processor) must be addressed for scalable implementation.~\cite{wang2024hardwareefficientquantumrandomaccess}. 

To overcome these challenges, we adopt several strategies tailored to the constraints of planar superconducting hardware. For layout optimization, we employ a triangular tiling topology that preserves binary-tree connectivity while minimizing inter-router wiring and avoiding loops, as shown in Fig.~\ref{fig 4}(b) and further discussed in the supplementary materials~\cite{supp_QRouter}. 
To reduce query latency, we adopt a similar circuit structure inspired by Refs.~\cite{chen2023efficient}, as illustrated in Fig.~\ref{fig 4}(d), with full circuit details provided in the supplementary materials~\cite{supp_QRouter}. This circuit follows a symmetric “$\vee$” structure to support both forward (top-down) and backward (bottom-up) routing—an essential feature for full QRAM operations. Under this architecture, layer-wise parallel activation of QRouters becomes possible in both directions, allowing routing information to propagate through the network with minimal latency and maintaining temporal symmetry.



As a proof-of-concept demonstration, we implemented a two-layer routing network using the above strategies and evaluated its routing performance and fidelity. This setup represents the minimal nontrivial configuration for a cascaded QRAM-compatible system and serves as an essential validation of the network's scalability.

First, we investigated address-path entanglement within the system. Using the setup illustrated in the inset of Fig.~\ref{fig 4}(c), we excited the top qubit and continuously varied the addresses (\(A^c_1\), \(A^c_2\)) of \(\rm C_1\) and \(\rm C_2\) through \(\theta\) adjustments in the multiple routing tests, while monitoring the data qubits \(\rm D_1\) and \(\rm D_2\). The experimental results in Fig.~\ref{fig 4}(c) depict the population distributions of \(\rm D_1\) and \(\rm D_2\) after routing. When \(\theta_1 = 0\) and \(\theta_2 = 1\), the population of \(\rm D_1\) reaches its maximum (\(P_1^{\rm max} = 93.96\%\)), consistent with the expected \(\sin^2(\theta_1)\cos^2(\theta_2)\). A similar trend is observed for \(\rm D_2\) (\(P_2^{\rm max} = 92.51\%\)) when \(\theta_1 = 0\) and \(\theta_2 = 0\), demonstrating the system's effective routing capability for information transfer.

We further conducted random access tests to evaluate the system's average fidelity. A randomized address sequence \(\{(A_1,A_1)\), \((A_2,A_2)\), \(\dots\), \((A_n,A_n)\), \((A_{n+1})\}\) was loaded, where each \(A_k\) comprised three random addresses \(\{a_1,a_2,a_3\}\). Each address sequence was initialized within the corresponding block and reset after routing, as shown in Fig.~\ref{fig 4}(e). The fidelity was determined by comparing the experimentally measured populations of the input qubits and four data qubits with the theoretically expected results, as illustrated in Fig.~\ref{fig 4}(f). 
The measured $M_{\rm RAT}$ at \(N=0\) were 90.02\% with the eraser scheme and 78.50\% without. The performance gradually declined with increasing routing depth due to factors such as addresses' two-photon transitions, imperfect operations, and data information decay.
Notably, the system employing the eraser scheme consistently outperformed its non-eraser counterpart, with fitted average fidelities of 82.40\% and 81.90\%, respectively, highlighting its potential for scalable QRAM implementation.

\section{\label{sec:level4}Discussion and conclusion}

In this study, we demonstrated the implementation of hardware-efficient, deterministic and complete-transmission QRouters on a superconducting quantum processor. Our work constitutes one of the first experimental realizations of a two-layer quantum routing network, enabling address-dependent, reversible routing with minimal hardware overhead. While previous efforts have largely remained at the theoretical level or focused on single-router demonstrations, our cascaded routing network showcases scalable, coherent routing compatible with grid-based superconducting hardware, marking a critical step toward practical QRAM architectures.

The coherent QRouters achieved reversible routing with a high transmission efficiency of 98\%, and the two-layer network maintained an overall efficiency of 93\%. In random access tests—a benchmark we introduce for scalable quantum routing systems—the single-layer and two-layer configurations achieved fidelities of 95.74\% and 82.40\%, respectively. 
These results validate theoretical observations that shallow circuits and post-selection can effectively suppress errors and improve fidelity in noisy intermediate-scale quantum devices~\cite{Preskill2018quantumcomputingin,lau2022nisq,Khalid2024Quantum}.

Through detailed characterization, we identified decoherence and control-induced leakage as dominant sources of infidelity, leading to asymmetric paths and occasional data misrouting~\cite{supp_QRouter}. These observations provide concrete targets for system-level optimization. Techniques such as post-selection, high-coherence fabrication, and waveform refinement offer practical pathways to mitigate such effects without introducing additional quantum overhead.

As the routing network scales further, new challenges emerge, including QRouter layout overlap, increased crosstalk~\cite{Suppression2019Mundada,Quantum2022Zhao,yang2024fast}, and error accumulation across multiple routing paths~\cite{supp_QRouter}. Addressing these issues will require advances in hardware layout, control sequencing, and short-range quantum interconnects~\cite{Rodrigo2021Modelling,Perfect2018Li,Perfect2005Christandl}. Nevertheless, our results establish both the architectural viability and the experimental feasibility of scalable QRAM-compatible routing, and provide a practical blueprint for extending coherent quantum data access on near-term superconducting platforms.
\\

\begin{acknowledgments}
This work has been supported by the National Key Research and Development Program of China (Grant No. 2023YFB4502500), and the National Natural Science Foundation of China (Grant No. 12404564).
This work is partially carried out at the USTC Center for Micro and Nanoscale Research and Fabrication. 
\end{acknowledgments}



\bibliography{QRouter}

\newpage


\newcommand{\beginsupplement}{%
	\setcounter{table}{0}%
	\renewcommand{\thetable}{S\arabic{table}}%
	\setcounter{figure}{0}%
	\renewcommand{\thefigure}{S\arabic{figure}}%
	\setcounter{equation}{1}%
	\renewcommand{\theequation}{S\arabic{equation}}%
	\setcounter{section}{0}%
	\renewcommand{\thesection}{\arabic{section}}%
        \setcounter{page}{1}%
        \renewcommand{\thepage}{S\arabic{page}}%
}

\let\oldaddcontentsline\addcontentsline
\renewcommand{\addcontentsline}[3]{}

\let\addcontentsline\oldaddcontentsline
\resetlinenumber
\clearpage
\onecolumngrid
\begin{center}
\textbf{\large Supplementary Information for ``Demonstrating Coherent Quantum Routers for Bucket-Brigade Quantum Random Access Memory on a Superconducting Processor''}
\end{center}


\beginsupplement
\renewcommand{\citenumfont}[1]{S#1}%
\renewcommand{\bibnumfmt}[1]{[S#1]}%

\NewDocumentCommand{\citesm}{>{\SplitList{,}} m }{%
  \def\temp{}%
  \ProcessList{#1}{\addSMprefix}%
  \expandafter\cite\expandafter{\temp}%
}
\newcommand{\addSMprefix}[1]{%
  \ifdefempty{\temp}%
    {\def\temp{SM_#1}}
    {\edef\temp{\temp,SM_#1}}%
}

\section{Device}{\label{Device}}

\begin{figure}[ht]
    \centering
    \includegraphics[width=1.\linewidth]{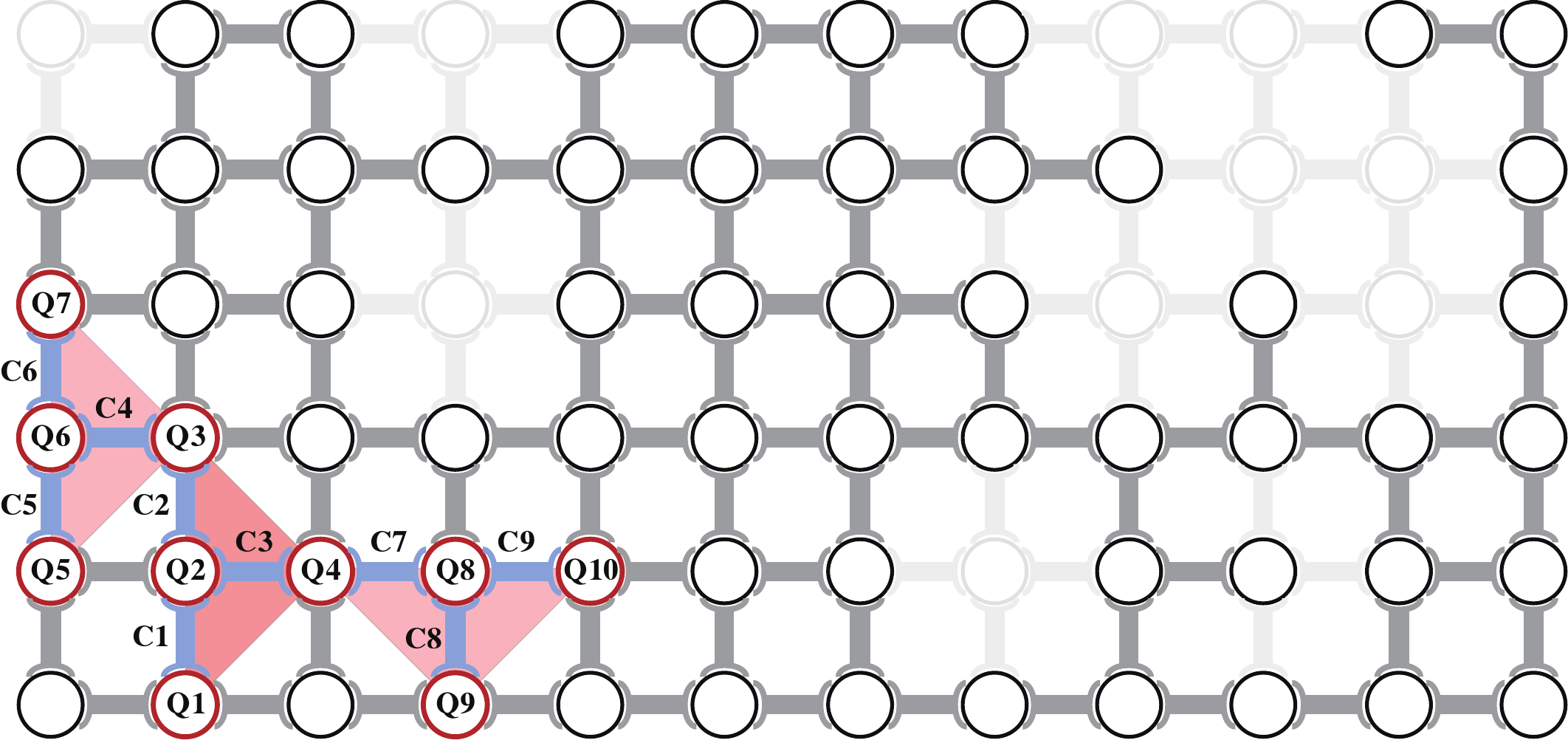}
    \caption{\justifying
    The topology of the superconducting processor \textit{Wukong} and qubits used in the experiment. Dark red regions indicate well-performing qubits and couplers, while light regions correspond to malfunctioning ones.
    Ten qubits (highlighted in red) were selected as the experimental platform, arranged in the form of three connected triangles. Each triangle functions as a quantum router.
    }
    \label{fig device}
\end{figure}

We conducted experiments on a superconducting quantum processor named \textit{Wukong}, comprising 72 transmon superconducting qubits (Q) and 126 transmon-type couplers (C) arranged in a grid-like topology. The topology of the chip is shown in Fig.~\ref{fig device}.
We selected ten of these qubits as our experimental platform, whose basic parameters and decoherence performance are listed in Table~\ref{tab device}.

\begin{table}[ht]
\centering
\caption{
Relevant qubit parameters and performances for the device under test. We list qubit transition frequencies from states $|i\rangle$ to $|j\rangle$ ($f_{ij}$) and anharmonicities $\alpha$,	while $f_{12}=f_{01}+\alpha$.
We also give qubit coherence parameters ($T_1$, $T^*_2$) and single-qubit gate fildeities ($F_{sq}$).
}
\begin{tabular}{ccccccccccc}
    \toprule[1pt]
    \textbf{Parameters} &  \textbf{Q1} & \textbf{Q2} & \textbf{Q3} & \textbf{Q4} & \textbf{Q5} & \textbf{Q6} & \textbf{Q7} & \textbf{Q8} & \textbf{Q9} & \textbf{Q10}\\ \midrule
    \textbf{$f_{01}\ \rm (GHz)$} & 4.269 & 4.622 & 4.163
    & 4.238 & 4.308 & 4.587 & 4.107 & 4.401 & 4.315 & 4.130\\ 
    \textbf{$\alpha \ \rm (MHz)$} & -246.6 & -188.4 & -242.4
    & -244.1 & -236.7 & -233.9 & -243.5 & -238.7 & -248.3 & -245.4\\ 	
    \midrule 
    \textbf{$T_1\ \rm (\mu\mathrm{s})$} & 21.543 & 18.729 & 19.465
    & 22.342 & 17.232
    & 23.474 & 7.728
    & 16.856 & 15.972
    & 22.531\\
    \textbf{$T^*_2\ \rm (\mu\mathrm{s})$} & 2.871 & 3.251 & 2.337
    & 2.105 & 2.178
    & 3.122 & 0.973
    & 1.842 & 1.619
    & 2.068\\ 			
    \midrule 
    \textbf{$F_{sq}\ \rm (\%)$} & 99.81 & 99.65 & 99.68
    & 99.37 & 99.68
    & 99.76 & 99.68
    & 99.59 & 99.78
    & 99.65\\		
    \bottomrule[1pt]
    \end{tabular}
    \captionsetup{justification=raggedright,singlelinecheck=false}
    \label{tab device}
\end{table}

\section{$\rm \sqrt{CZ}$ Gate optimization based on Floquet sequences}{\label{Floquet}}

The \(\sqrt{\text{CZ}}\) gate is a key component in constructing the quantum router demonstrated in this work. This section presents the theoretical analysis and experimental optimization of the \(\sqrt{\text{CZ}}\) gate, along with relevant statistical data. We began by studying the exchange dynamics between states $|11\rangle$ and $|02\rangle$ ($\mathrm{iSWAP}_{02\leftrightarrow 11}$ operation) to determine suitable Floquet sequences and optimization targets~\citesm{arute2020observation,neill2021accurately,gross2024characterizing}. 
The mathematical representation of the $\mathrm{iSWAP}_{02\leftrightarrow 11}$ operation in the  $\{|11\rangle,|02\rangle\}$ basis is given by:
\begin{gather}
\mathrm{iSWAP}_{02\leftrightarrow 11}(\vartheta,\eta)= 
    \begin{pmatrix}
    e^{i\eta}\cos{(\vartheta/2)} & -     i\sin{(\vartheta/2)}  \\
        -i\sin{(\vartheta/2)} & e^{-i\eta}\cos{(\vartheta/2)}
    \end{pmatrix}, \label{CP} \\
    \sqrt{\text{CZ}} = \mathrm{iSWAP}_{02\leftrightarrow 11}(\pi,\eta)=
    \begin{pmatrix}
        0 & -i  \\
        -i & 0
    \end{pmatrix}, 
\end{gather}
where $\vartheta$ and $\eta$ denote the Rabi angle and the controlled phase acquired during the operation~\citesm{rol2019fast}. Setting \(\vartheta = \pi\) realizes the ideal \(\sqrt{\text{CZ}}\), which fully exchanges the populations of states \(|11\rangle\) and \(|02\rangle\). Floquet theory can be used to analyze the dynamics of repeated \(\sqrt{\text{CZ}}\) gates. The evolution under 
$N$ applications of the composite gate $({Z_{qq}\cdot \sqrt{\text{CZ}}}\cdot Z_{qq})^{N}$ is described by:
\begin{gather}
   ({Z_{qq}\cdot \sqrt{\text{CZ}}}\cdot Z_{qq})^{N}=e^{-iN\Omega\sigma_n}=I\cos N\Omega-i\sigma_n\sin N\Omega,
\end{gather}
where 
\begin{gather}	\Omega=\arccos(\cos(\vartheta/2)\cos(\zeta/2-\eta)), \\
    Z_{qq}
    = 
    \begin{pmatrix}
    1 & 0  \\
    0 & e^{i\zeta/2}
    \end{pmatrix},
    \sigma_n = 
    \begin{pmatrix}
    \cos\alpha & \sin\alpha \\
    \sin\alpha & -\cos\alpha
    \end{pmatrix},
    \alpha=\arctan(\frac{\tan(\vartheta/2)}{\sin(\eta/2-\zeta)}).
    \end{gather}

For $\vartheta=\pi$, we have:
\begin{gather}
    \Omega=\pi/2, \ \ \
    \alpha=\pi/2,\\
    ({Z_{qq}\cdot \sqrt{\text{CZ}}}\cdot Z_{qq})^N=
    \begin{pmatrix}
    e^{i(\eta+\zeta/2)}\cos{(N\pi/2)} & -i\sin{(N\pi/2)}  \\
    -i\sin{(N\pi/2)} & e^{-i(\eta+\zeta/2)}\cos{(N\pi/2)}
    \end{pmatrix}.
\end{gather}
As a result, the population of state $|11\rangle$ ($P^{|11\rangle}_{2i}, i=0,1,2...$) reaches 1 after an even number of gates, while the population of state $|02\rangle$ ($P^{|02\rangle}_{2i+1}$) reaches 1 after an odd number of gates.
Under the application of \(2m\) successive \(\sqrt{\text{CZ}}\) gates, we define the average population of these two states as
cost $C^m_l = \sum^{m-1}_{i=0}{(P^{|02\rangle}_{2i+1}+P^{|11\rangle}_{2i})}/2m$,
which characterizes leakage and decoherence. For ideal operations, $C^m_l$ reaches 100\%.

\begin{figure}[ht]
    \centering
    \includegraphics[width=0.65\linewidth]{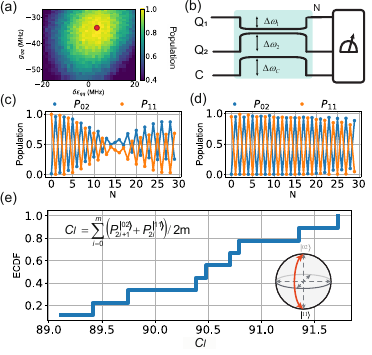}
    \caption{\justifying
    (a) Landscapes of $P_{02}$ as a function of detune ($\delta \varepsilon_{qq}=\varepsilon_{|11\rangle}-\varepsilon_{|02\rangle}$) and effective coupling strength ($g_{qq}$).
    The operation time is 25 ns.
    (b) show the schematic illustration of the setup for optimization. 
    (c) and (d) show the population evolution before and after optimization, respectively.
    (e) Distribution of $C^{15}_l$ of nine $\sqrt{\rm CZ}$ gates, with 5000 shots per experiment.
    The expected value of $\sqrt{\rm CZ}$ is 100\%.
    }
    \label{fig floquetCZ}
\end{figure}

We first adjusted the waveform parameters of the $\sqrt{\rm CZ}$ gate (Fig.~\ref{fig floquetCZ}(a)),
using a peak-finding algorithm to achieve full population exchange between states $|11\rangle$ and $|02\rangle$.
Next, we employed an $N$-repeat gate circuit along with a Nelder-Mead (NM) algorithm to optimize waveform parameters ($\Delta\omega_1,\Delta\omega_2,\Delta\omega_C$) under a fixed duration ($t = 25~{\rm ns}$) in parallel, 
as illustrated in Fig.~\ref{fig floquetCZ}(b).  
The cost function was then defined as $C^{m=15}_l=\sum^{14}_{i=0}{(P^{|02\rangle}_{2i+1}+P^{|11\rangle}_{2i})}/30$, aiming to minimize unwanted rotation and the influence of spurious two-level systems.  
Figures~\ref{fig floquetCZ}(c) and (d) compare the gate performance before and after optimization. 
After optimization, the oscillation envelope associated with leakage was significantly suppressed, indicating reduced leakage. Similar optimization procedures were performed for all qubit pairs, with the results summarized in Fig.~\ref{fig floquetCZ}(e).
We observed that even when using optimized parameters and a flat-Gaussian waveform, some \(\sqrt{\text{CZ}}\)  exhibited residual leakage. A likely technical source of this issue is waveform distortion, which hampers perfect population transfer between states \(|11\rangle\) and \(|02\rangle\).
This effect can be mitigated using more precise digital filtering techniques, such as finite impulse response (FIR) or infinite impulse response (IIR) filters~\citesm{butscher2018shaping,johnson2011controlling,chen2018metrology,yan2019strongly}.  

\section{Controlled SWAP operation based on Transition Composite Gate scheme}{\label{TCG scheme}}

\begin{figure}[ht]
\centering
\includegraphics[width=0.65\linewidth]{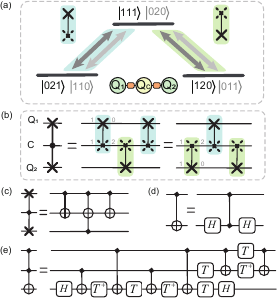}
\caption{\justifying
CSWAP operation based on the TCG scheme.
(a) Transition pathways of our CSWAP operation. Two parallel transition paths ($|110\rangle \leftrightarrow |020\rangle \leftrightarrow |011\rangle$, $|120\rangle \leftrightarrow |111\rangle \leftrightarrow
|021\rangle$) enable CSWAP conditional on the control qutrit states $(|1\rangle,|2\rangle)$.
(b) Two equivalent combinatorial implementations of the CSWAP operation, each requiring only three two-qubit gates.
(c)-(e) Standard CSWAP construction based on Clifford decompositions, requiring 20 single-qubit and 16 two-qubit gates.
}
\label{fig CSWAPde}
\end{figure}

This section presents a comprehensive overview of the theoretical framework and experimental results for the controlled SWAP (CSWAP) operation developed in this study~\citesm{chapman2023high,nielsen2010quantum}. 
In previous work, we proposed the Transition Composite Gate (TCG) scheme, demonstrating that controlled operations can be implemented with shallow circuit depths by utilizing high-energy levels as auxiliary energy levels through  transition pathway engineering~\citesm{zhang2024realization}. 
Under this idea, the TCG-based CSWAP gate requires only three \(\sqrt{\text{CZ}}\) gates, 
with the corresponding transition pathway shown in Fig.~\ref{fig CSWAPde}(a).
Analysis of transition pathways reveals that this operation exhibits multiple-flag switching capabilities. 
This diode remains in the off state when the control unit is in state \(|0\rangle\).
Conversely, it conducts when the control unit is in states \(|1\rangle\) or \(|2\rangle\). 
This can be described in the Hilbert subspace spanned by the basis states \(\{|011\rangle, |020\rangle, |110\rangle, |111\rangle, |021\rangle, |120\rangle\}\):

\begin{equation}\label{eq:cswap_definition}
 \begin{aligned}
    {\rm CSWAP} &= \sqrt{\rm CZ}_{\rm Q_1Q_C} \otimes I_{\rm Q_2}\cdot I_{\rm Q_1} \otimes \sqrt{\rm CZ}_{\rm Q_CQ_2} \cdot \sqrt{\rm CZ}_{\rm Q_1Q_C} \otimes I_{\rm Q_2} \\ 		
    &= I_{\rm Q_1} \otimes \sqrt{\rm CZ}_{\rm Q_CQ_2}\cdot \sqrt{\rm CZ}_{\rm Q_1Q_C} \otimes I_{\rm Q_2} \cdot I_{\rm Q_1} \otimes \sqrt{\rm CZ}_{\rm Q_CQ_2} \\ 
    &= (-1)\cdot
    \begin{pmatrix}
        0 & 0 & 1 & 0 & 0 & 0 \\
        0 & 1 & 0 & 0 & 0 & 0 \\
        1 & 0 & 0 & 0 & 0 & 0 \\
        0 & 0 & 0 & 1 & 0 & 0 \\
        0 & 0 & 0 & 0 & 0 & 1 \\
        0 & 0 & 0 & 0 & 1 & 0 \\
    \end{pmatrix},
 \end{aligned}
\end{equation}
where
\begin{equation}
    \sqrt{\rm CZ}_{q_1q_c} \otimes I_{q2} 
    = 
    \begin{pmatrix}
        1 & 0 & 0 & 0 & 0 & 0 \\
        0 & 0 & -i & 0 & 0 & 0 \\
        0 & -i & 0 & 0 & 0 & 0 \\
        0 & 0 & 0 & 0 & -i & 0 \\
        0 & 0 & 0 & -i & 0 & 0 \\
        0 & 0 & 0 & 0 & 0 & 1 \\
    \end{pmatrix},
\end{equation}
and
\begin{equation}
    I_{\rm Q1} \otimes \sqrt{\rm CZ}_{\rm Q_CQ_2}
    = 
    \begin{pmatrix}
        0 & -i & 0 & 0 & 0 & 0 \\
        -i & 0 & 0 & 0 & 0 & 0 \\
        0 & 0 & 1 & 0 & 0 & 0 \\
        0 & 0 & 0 & 0 & 0 & -i \\
        0 & 0 & 0 & 0 &1 & 0 \\
        0 & 0 & 0 & -i & 0 & 0 \\
    \end{pmatrix},
\end{equation}
and $I$ denotes the identity operation. Here, $\rm Q_1$ and $ \rm Q_2$ are data qubits, and $\rm Q_C$ is the control qutrit.

The CSWAP operation can be implemented in two equivalent ways, each corresponding to a distinct yet symmetric transition paths, as shown in Fig.~\ref{fig CSWAPde}(b).
In both cases, the operation matrix is block-diagonal, and the swap occurs within two decoupled subspaces of the Hilbert space. Compared to the conventional Clifford-based decomposition—which requires circuit depths up to 12  (Fig.~\ref{fig CSWAPde}(c)-(e))—this scheme provides a substantial reduction in complexity. Importantly, the absence of any additional phase factors in Eq.~\eqref{eq:cswap_definition} eliminates the need for further phase corrections.

We acknowledge that similar work has been conducted recently in this field~\citesm{li2024hardware}. However, our analysis offers a distinct perspective through the TCG framework, proposing the utilization of higher energy levels as switches. 
Furthermore, the short-path TCG (SP-TCG) reveals shorter pathways for unidirectional transmission, offering significant insights for short-range quantum information transfer~\citesm{zhang2024realization}. 
For unidirectional information transfer from $\mathrm{Q}_1$ to $\mathrm{Q}_2$, only two \(\sqrt{\text{CZ}}\) gates are required, as demonstrated in the following part. The shorter pathway provides advantages for large-scale quantum information processing and quantum computing. 	Additionally, 
it retains digital benefits, 
enabling rapid initialization of quantum systems and enhancing scalability in extensive architectures.

\section{Quantum router}{\label{address-path}}
\subsection{Quantum router and address-path entanglement validation}{\label{address-path}}

This section presents the theoretical framework and experimental validation of address–path entanglement in the QRouter, using the non-eraser scheme ($|L\rangle = |0\rangle, |R\rangle = |1\rangle$) as a representative case for simplicity.
The QRouter is composed of two CSWAP gates and a sequence of single-qubit operations. The validation process is divided into two parts, examining the control qutrit
$\rm Q_C$'s state $|\psi_c\rangle=\cos\theta|0\rangle +e^{i\phi}\sin\theta|1\rangle$ 
in terms of the parameters 
\(\theta\) and \(\phi\), respectively.

The parameter \(\theta\) overns the routing ratio, determining the probability amplitude for directing the input qubit (\(\mathrm{Q}_{\rm I}\)) to the left path qubit (\(\mathrm{Q}_{\rm L}\)) or to the right path qubit (\(\mathrm{Q}_{\rm R}\)).
This can be described in the Hilbert space spanned by \(\{|0010\rangle, |1000\rangle, |0101\rangle, |1100\rangle\}\): 
\begin{gather} 
    U_{\rm Q_IQ_CQ_LQ_R} = {\rm CSWAP}_{\rm Q_IQ_CQ_L}\otimes{I}_{\rm Q_R} \cdot  {I}_{\rm Q_I}\otimes{X}_{\rm Q_C}\otimes{I}_{\rm Q_I}\otimes{I}_{\rm Q_R} \cdot {I}_{\rm Q_L}\otimes{\rm CSWAP}_{\rm Q_IQ_CQ_R}
    \cdot  {I}_{\rm Q_I}\otimes{X}_{\rm Q_C}\otimes{I}_{\rm Q_L}\otimes{I}_{\rm Q_R} \label{eq:QR}\\
    = 
    \begin{pmatrix}
        0 & -1 & 0 & 0 \\
        -1 & 0 & 0 & 0 \\
        0 & 0 & 0 & -1 \\
        0 & 0 & -1 & 0
    \end{pmatrix}, \\
    S= e^{i\tau}
    \begin{pmatrix}
        0, & \sin\theta, & 0, & e^{i\phi}\cos\theta 
    \end{pmatrix}^T, \\
    S' = US = e^{i\tau}
    \begin{pmatrix}
        \sin\theta, & 0, & e^{i\phi}\cos\theta, & 0 
    \end{pmatrix}^T, \\
    P_{|0010\rangle} = \sin^2\theta,\ \  
    P_{|0101\rangle} = \cos^2\theta,
\end{gather}
where $S$ and $S'$ are initial and final states, respectively. And $\tau$ is an adjustable global phase.	

\begin{figure}[ht]
\centering
    \includegraphics[width=0.7\linewidth]{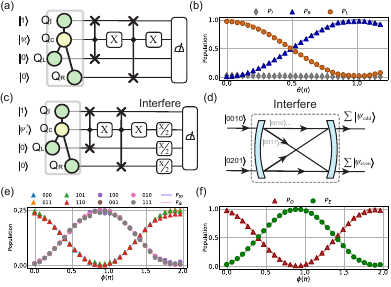}
    \caption{\justifying
        Verification of address-path entanglement in the QRouter. 
        (a) and (c) show experimental setups for verifying \(\theta\) and \(\phi\) in the address, respectively. 
        (b) Experimental results of varying $\theta$.
        The routing ratio changes with different values of $\theta$,
        while $\phi$ has no effect on it.
        (d) Interference principle diagram.
        After $X_{\pi/2}$ operations, a cat state ($|0010\rangle+e^{i\phi}|0101\rangle$) undergoes interference between states with even and odd excitation numbers, resulting in distinct population distributions.  
        (e) Experimental interference results of changing $\phi$.
        The population of each odd-excitation state
        satisfies \((1-\sin{\phi})/8\), 
        while even-excitation states
        exhibit complementary behavior (\((1+\sin{\phi})/8\)).
        (f) The total population of odd-excitation and even-excitation states graph corresponding to (e).
    }
    \label{fig inter}
\end{figure}

We experimentally vary \(\theta\) using the setup in Fig.~\ref{fig inter}(a),
starting from the initial state	
$|100\rangle_{\rm Q_IQ_LQ_R} \otimes |\psi_{c}\rangle_{\theta,\phi}=\sin\theta |1000\rangle_{\rm Q_IQ_CQ_LQ_R} + e^{i\phi}\cos\theta |1100\rangle_{\rm Q_IQ_CQ_LQ_R}$.
The measured populations $P_{|0010\rangle}$ and $P_{|0101\rangle}$ follow the expected cosine dependence on $\theta$, as shown in Fig.~\ref{fig inter}(b), validating deterministic and complete-transmission quantum routing behavior.

In contrast, the parameter \(\phi\) governs the phase relationship in the address–path entangled state, but it does not affect the routing ratio, rendering its influence less direct.
By initializing the system in the superposition  $|100\rangle_{\rm Q_IQ_LQ_R} \otimes |\psi_{c}\rangle_{\pi/4,\phi}=(|1000\rangle_{\rm Q_IQ_CQ_LQ_R} + e^{i\phi}|1100\rangle_{\rm Q_IQ_CQ_LQ_R})/\sqrt{2}$  and inserting an interference layer (Fig.~\ref{fig inter}(c)), we generate a cat state subject to coherent interference (Fig.~\ref{fig inter}(d)).
The relative phase $\phi$ changes the phase difference between the entangled states generated under coherent control,
which leads to various interference processes.
Consequently, phase effects can be observed by measuring populations of odd-excitation subspaces ($|0001\rangle, |0010\rangle,	|0100\rangle,|0111\rangle$) or even-excitation subspaces ($|0000\rangle, |0011\rangle,
|0101\rangle,|0110\rangle$), 
whose parity is determined by the number of qubits in the non-zero state.
This can be described in \(\{|0000\rangle, |0001\rangle, |0010\rangle,|0011\rangle,\) \(|0100\rangle,|0101\rangle,|0110\rangle,|0111\rangle\}\) as follows:

\begin{align}
    S'_{+} &= US_{+}=		
    \begin{pmatrix}
        0 & 0 & 1 & 0 & 0 & e^{i\phi} & 0 & 0
    \end{pmatrix}^T /\sqrt{2}, \\
    S''_{+} &= I \otimes X_{\pi/2} \otimes X_{\pi/2} \otimes X_{\pi/2} \cdot S'_{+} \\
    &=\frac{1}{4}	
    \begin{pmatrix}
        -i-e^{i\phi} \\
        -1-ie^{i\phi} \\
        1+ie^{i\phi} \\
        -i-e^{i\phi} \\
        -1-ie^{i\phi} \\
        i+e^{i\phi} \\
        -i-e^{i\phi} \\
        -1-ie^{i\phi}
    \end{pmatrix},\\
    P &=S''_{+}{S''_{+}}^* \\
    &= \frac{1}{8}	
    \begin{pmatrix}
        1+\sin{\phi} \\
        1-\sin{\phi} \\
        1-\sin{\phi} \\
        1+\sin{\phi} \\
        1-\sin{\phi} \\
        1+\sin{\phi} \\
        1+\sin{\phi} \\
        1-\sin{\phi}
    \end{pmatrix},
\end{align}
where $S_+$, $S'_+$, and $S''_+$ are the  initial states, middle states before interference, and final states after interference,
respectively.

As a result, the total population of odd-excitation states, denoted $P_{O}$, follows \((1-\sin{\phi})/2\), 
while that of even-excitation states
$P_{E}$ follows \((1+\sin{\phi})/2\).
A fixed initial phase offset is implicitly included in $\phi$.
Experimental data in Figs.~\ref{fig inter}(e) and (f) match theoretical predictions $P_{tO}$ and $P_{tE}$, confirming that the QRouter realizes coherent address–path entanglement via controlled interference.

\subsection{Quantum state tomography results for quantum routers}{\label{QST}}

\begin{figure}[ht]
    \centering
    \includegraphics[width=0.5\linewidth]{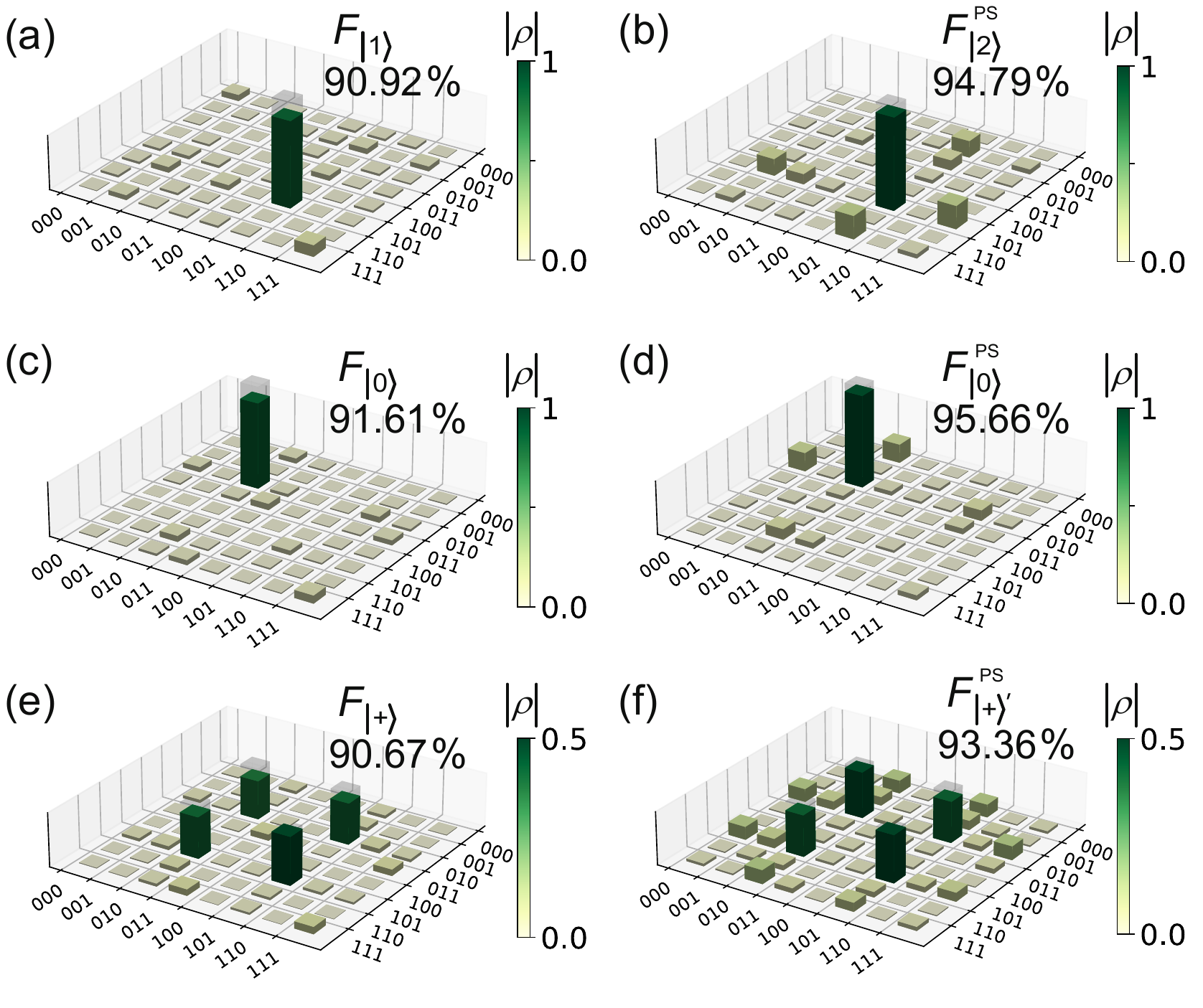}
    \caption{\justifying
        Quantum state tomography (QST) results for QRouters under the non-eraser scheme (left) and eraser scheme (right).
        (a), (c) and (e) show density matrices under addresses $|1\rangle$, $|0\rangle$ and $|+\rangle=(|0\rangle+|1\rangle)/\sqrt 2$,
        respectively.
        (b), (d) and (f) show density matrices after post-selection under addresses $|2\rangle$, $|0\rangle$, and $|+'\rangle=(|0\rangle+|2\rangle)/\sqrt 2$,
        respectively.
        Each datapoint  undergoes 10 QST experiments with 10000 shots.
    }
    \label{fig QST}
\end{figure}

This section presents quantum state tomography (QST) results for the QRouter under various address configurations,
with and without eraser scheme, as illustrated in Fig.~\ref{fig QST}.
Eraser-detection improves QST fidelity by approximately 3\%-5\%, though further enhancement remains possible. 
Residual errors may stem from decoherence, as well as from imperfect state preparation and \(\sqrt{\mathrm{CZ}}\) operations.

\section{Quantum routing network towards Quantum random access memory}{\label{QRAM}}
\subsection{Routing system layout design}{\label{layout}}

We present a practical layout strategy for implementing a quantum routing system on a lattice-based quantum processor.
Realizing a QRAM prototype requires the systematic arrangement and interconnection of QRouters into a binary-tree routing network—a task that poses considerable engineering challenges.
As the scale of the system increases, the exponential growth in the number of quantum units makes compact arrangement on a 2D grid topology increasingly difficult, especially when considering unavoidable defects in qubits or couplers~\citesm{wang2024hardwareefficientquantumrandomaccess}. 

To address this issue, we propose a densely packed triangular scheme (PTS).
In this model, each QRouter is represented as an isosceles triangle comprising four qubits positioned on a 2D grid.
The overall QRAM network forms a cascade of such triangles, emulating a binary tree.
To optimize qubit utilization while minimizing the impact of defective qubits or couplers, triangles are densely packed in accordance with two key design principles: i) triangles are connected at vertices in pairs (QRouter connections form a tree structure), ii) triangles do not enclose a region (QRouters do not overlap). 

\begin{figure}[ht]
\centering
    \includegraphics[width=0.9\linewidth]{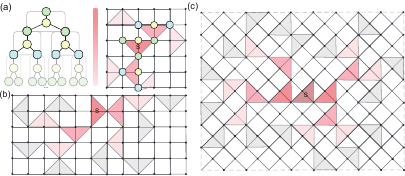}
    \caption{\justifying
    The layout of the routing network for QRAM on the grid-based quantum processor. 
    (a) QRouters are arranged in an isosceles triangular pattern within a regular 2D grid topology. Through optimization constraints and layout algorithms, we achieved an optimal arrangement of the processor. The figure illustrates a two-layer quantum routing network layout, 
    where the color gradient from dark red to light red indicates increasing layers. The central qubit (yellow) in each QRouter serves as the address qubit. while the other qubits (green and blue) function as data qubits.
    Considering hardware defects and decoherence limitations, a portion of the layout is truncated for the practical implementation of a two-layer system.  
    (b) An example of a four-layer network layout on a defect-free $12\times6$ processor.
    (c) An example of the densest five-layer system layout on a defect-free processor.
        }
    \label{fig layout}
\end{figure}	

We developed a layout algorithm that initiates from a seed triangle and iteratively “grows” triangular links in topologically favorable regions.
By scanning possible seed locations, we identified the most compact and defect-tolerant configuration.
The selected layout, shown in Fig.~\ref{fig layout}, was chosen based on a holistic evaluation of qubit coherence, crosstalk, and hardware constraints.
We further demonstrate the optimal packing for the ideal \textit{Wukong} processor with a $12 \times 6$ topology. 
Interestingly, the optimal configuration places the seed triangle near the processor’s center, suggesting a quasi-circular layout for maximizing packing efficiency..

Our findings indicate that this layout can accommodate up to a five-layer system.
Scaling beyond this limit inevitably introduces overlaps between QRouters, unless supplemented by short-distance transmission techniques.
This observation highlights the necessity for hardware-aware designs to enable further expansion.
Nonetheless, our packing strategy provides a scalable architectural framework and valuable design insights for the realization of large-scale quantum routing networks.

\subsection{Quantum routing system verification}{\label{system routing}}

\begin{figure}[ht]
    \centering
    \includegraphics[width=1.0\linewidth]{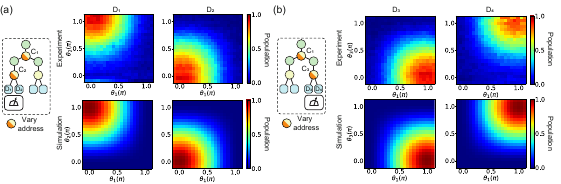}
    \caption{\justifying
        Experimental verification of two-layer system routing performance.
        (a) and (b) show the routing behavior for two independent branches, involving data qubit pairs \{$\mathrm{D}_1$, $\mathrm{D}_2$\} and \{$\mathrm{D}_3$, $\mathrm{D}_4$\} respectively.
        Population distributions are measured as a function of address-control parameters $\theta_1$ and $\theta_2$.
    }
    \label{fig system routing}
\end{figure}

We experimentally verified the complete routing functionality of our two-layer quantum routing network.
Figure~\ref{fig system routing} illustrates the experimental configurations and corresponding results.
In the first experiment, we continuously varied the control addresses ($A^c_1$,$A^c_2$) of control nodes $\mathrm{C}_1$ and $\mathrm{C}_2$ by adjusting the parameters $\theta_1$ and $\theta_2$, while monitoring the population dynamics of the target data qubits $\mathrm{D}_1$ and $\mathrm{D}_2$.
The resulting population landscapes, shown in Fig.~\ref{fig system routing}(a), reveal that when
$\theta_1 = 0$ and $\theta_2 = \pi$, the population of $\mathrm{D}_1$ reaches a maximum of $P_1^{\rm max} = 93.96\%$, which agrees well with the theoretical expectation $\sin^2(\theta_1)\cos^2(\theta_2)$.
Similarly, for $\theta_1 = 0$ and $\theta_2 = 0$,
we observe a maximum population of $P_2^{\rm max} = 92.51\%$ in $\mathrm{D}_2$ 
confirming reliable and selective quantum routing behavior.
Numerical simulations further corroborate the experimental results, reinforcing the effectiveness of the routing mechanism.
A second set of experiments was performed on an independent routing branch, targeting data qubits $\mathrm{D}_3$ and $\mathrm{D}_4$, as shown in Fig.~\ref{fig system routing}(b).
We obtained maximum populations of $P_3^{\rm max} = 92.23 \%$ and $P_4^{\rm max} = 88.38\%$ (Fig.~\ref{fig system routing}(b)), respectively.
The slightly reduced population observed in $\mathrm{D}_4$ is primarily attributed to the limited coherence times of its associated physical qubit, $\mathrm{Q}_7$, with $T_1 = 7.728 ~\mu\mathrm{s}$ and $T_2^* = 0.973~\mu\mathrm{s}$.

\subsection{Structure and quantum circuits of quantum random access memory }{\label{quantum circuits}}

This section presents the structural design and quantum circuit implementation of quantum random access memory (QRAM). 

\begin{figure}[ht]
    \centering
    \includegraphics[width=0.8\linewidth]{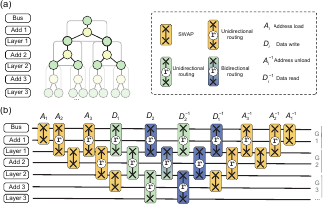}
    \caption{\justifying
        (a) Schematic structure of the QRAM. Qubits are organized into a binary tree: green circles represent data qubits, yellow circles indicate control qubits, and gray boxes correspond to QRouter modules.
        (b) Quantum circuit of QRAM operations. The process begins with address loading, followed by bidirectional data transfer along the routed path. The desired data is then read out, and the system is restored by address unloading via inverse operations.
    }
    \label{fig:qram_circuit}
\end{figure}

The overall QRAM structure, depicted in Fig.~\ref{fig:qram_circuit}(a), adopts a binary tree architecture, where qubits are hierarchically divided into a input bus, address layers, and data layers.
At the top of the tree resides the bus, which orchestrates both address and data transmission.
Each gray box denotes a QRouter module, whose intermediate qubits serve as control nodes, thereby forming the address layer.
The remaining qubits constitute the data layer, responsible for routing and storing information; the lowest-level data qubits are typically connected to classical memory resources~\citesm{nielsen2010quantum,paler2020parallelizing,weiss2024qram,jaques2023qram,shi2024error,chen2021scalable}. 
 
To address query latency and enhance efficiency, we implemented the quantum circuits illustrated in Fig.~\ref{fig:qram_circuit}(b).  
Orange blocks represent address load/unload operations, while green and blue blocks correspond to data read/write operations.
The address is sequentially loaded into control qubits through SWAP and QRouter operations, and the corresponding data is routed along the defined path.
Upon retrieval, both address and data are reversed through the same route, completing the readout and reset process.
To maintain uniform query latency across different address paths, all SWAP and QRouter gates are applied in a fixed sequential order.
Moreover, within each parity group, either odd-indexed (G\(_{2k+1},\ k \in {\mathcal{R}}\)) or even-indexed (G\(_{2k}\)) groups, operations can be executed in parallel.
This controlled parallelism is enabled by the modular structure of the digital QRouter components.

We observed that many tasks involve unidirectional input or output operations only at the beginning and end of the QRAM process to load and unload addresses.
This observation allows us to apply short-path TCG (SPTCG) optimization techniques, as detailed in the following sections. 
By engineering asymmetric routing paths, we reduce circuit depth for unidirectional access patterns.
Moreover, under the TCG framework, SWAP gates can be decomposed into a combination of a $\rm \sqrt{CZ}$ gate and single-qubit rotations, effectively replacing the standard three-CNOT implementation. This substitution significantly mitigates cumulative gate error, particularly from two-qubit operations, which typically dominate quantum error rates~\citesm{zhang2024realization}.

\section{Simulation and analysis }{\label{SA}}
\subsection{Simulation results of eraser-detection}{\label{PS}}

This section presents a theoretical and simulation-based analysis of the effects of decoherence and eraser-detection with post-selection (PS).  
Decoherence refers to the interaction between a quantum system and its environment, which results in a probabilistic transition from higher to lower energy states.
In a qubit-based system, the decay rate from the excited state $|1\rangle$ to the ground state $|0\rangle$ is denoted by \(\Gamma_{10}\), and the corresponding population dynamics are given by:
\begin{align}
    \frac{dA_{|1\rangle}}{dt}&=-\Gamma_{10} A_{|1\rangle},\\
    \frac{dA_{|0\rangle}}{dt}&=\Gamma_{10} A_{|1\rangle},
\end{align}
with initial condition $A_{|1\rangle}(t=0)=1$, yielding:
\begin{align}
    A_{|1\rangle}(t)&=e^{-\Gamma_{10} t},\\
    A_{|0\rangle}(t)&=1-e^{-\Gamma_{10} t},
\end{align}
where $A_{|i\rangle}$ represents the probability amplitude of state $|i\rangle$.
The rate $\Gamma_{10}$ is the inverse of the decoherence time ($T_{10}$) as $\Gamma_{10} = 1/T_{10}$.

In realistic transmon systems, higher excited states such as state $|2\rangle$ exist, introducing an additional decay pathway characterized by $\Gamma_{21} = 1/T_{21}$,
 representing the relaxation from the state $|2\rangle$ to the state $|1\rangle$. 
In contrast, direct transitions from state $|2\rangle$ to state $|0\rangle$ are second-order processes with a significantly lower probability. The full dynamics of the three-level system are:
\begin{equation}
\begin{aligned}
    \frac{dA_{|2\rangle}}{dt}&=-\Gamma_{21} A_{|2\rangle},\\
    \frac{dA_{|1\rangle}}{dt}&=\Gamma_{21} A_{|2\rangle}-\Gamma_{10} A_{|1\rangle},\\
    \frac{dA_{|0\rangle}}{dt}&=\Gamma_{10} A_{|1\rangle},
\end{aligned}
\end{equation}
with $A_{|2\rangle}(t=0)=1$, yielding:
\begin{align}
   A_{|2\rangle}(t)&=e^{-\Gamma_{21} t},\\
   A_{|1\rangle}(t)&=\frac{\Gamma_{21} (e^{-\Gamma_{10} t}-e^{-\Gamma_{21} t})}{\Gamma_{21}-\Gamma_{10}},\\
    A_{|0\rangle}(t)&=\frac{\Gamma_{10} e^{-\Gamma_{21} t}-\Gamma_{21} e^{-\Gamma_{10} t}}{\Gamma_{21}-\Gamma_{10}}+1,
\end{align}

We consider states $|2\rangle$ and $|0\rangle$ as qubitized states, using post-selection techniques to detect and filter out the state $|1\rangle$.  
As a result, decoherence in the system manifests as the decay from the state $|2\rangle$ to the state $|0\rangle$, which is a second-order process.  
The renormalized populations of the target state $|2\rangle$ can subsequently be described as follows:
\begin{align}
    A'_{|0\rangle}(t)&=\frac{A_{|0\rangle}}{A_{|0\rangle}+A_{|2\rangle}},\\
    A'_{|2\rangle}(t)&=\frac{A_{|2\rangle}}{A_{|0\rangle}+A_{|2\rangle}}.
\end{align}

By encoding logical states in more widely separated energy levels and increasing the order of transitions, post-selection enhances robustness to both decoherence and thermal excitations (upward transitions from states $|0\rangle$ to $|1\rangle$ to $|2\rangle$). 

\begin{figure}[ht]
    \centering
    \includegraphics[width=0.6\linewidth]{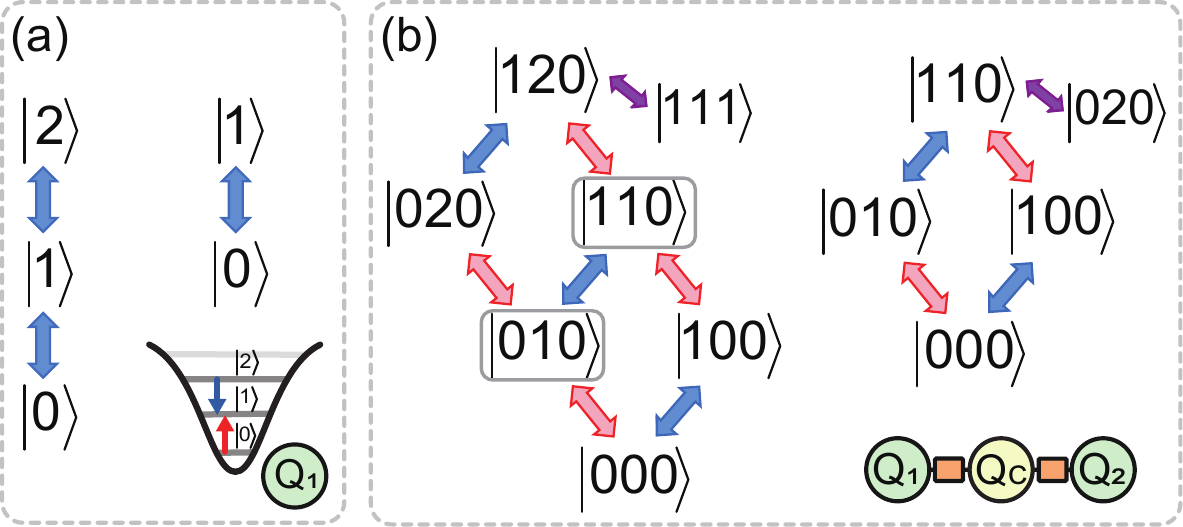}
    \caption{\justifying
        Energy level diagram.
        (a) Energy level diagram for a single qutrit. Post-selection can resolve the evolution of the state $|1\rangle$ caused by thermal excitation and decoherence.  
        (b) Energy level diagram for a three-qubit system. 
        Blue arrows indicate data mutation, red arrows denote address mutation, and purple arrows represent leakage caused by $\sqrt{\rm CZ}$ operations. 
    }
    \label{fig deconh}
\end{figure}

Our quantum routing system is constructed based on the TCG scheme, which supports diode functionality with the switch $\{|0\rangle,|2\rangle\}$ (eraser scheme) or  $\{|0\rangle,|1\rangle\}$ (non-eraser scheme).  
Since the system consists of stacked CSWAP operations, we model decoherence using a minimal three-qubit unit composed of two data qubits \(q_{1,2}\) and a control qubit \(q_c\), assuming identical \(\Gamma_{10}\) and \(\Gamma_{21}\) for all.  
We examine the initial states $|120\rangle$ and $|110\rangle$, with energy level diagrams shown in Fig.~\ref{fig deconh}.  

When using the non-eraser scheme, both data and control qubits decay via first-order processes, leading to full relaxation into the state $|000\rangle$.  
In contrast, under the eraser scheme, the first-order decay of the control qubit to state $|1\rangle$ can be filtered via post-selection, reducing the impact to second-order.
However, first-order decoherence in data qubits remains, highlighting the need for full second-order encoding in future designs.
The system dynamics for the initial state $|120\rangle$ under decoherence are:
\begin{align}
    \frac{dA_{|120\rangle}}{dt}&=-\Gamma_{10} A_{|120\rangle}-\Gamma_{21} A_{|120\rangle},\\
    \frac{dA_{|020\rangle}}{dt}&=\Gamma_{10} A_{|120\rangle}-\Gamma_{21} A_{|020\rangle},\\
    \frac{dA_{|110\rangle}}{dt}&=\Gamma_{21} A_{|120\rangle}-2\Gamma_{10} A_{|110\rangle},\\
    \frac{dA_{|010\rangle}}{dt}&=\Gamma_{21} A_{|020\rangle}+\Gamma_{10} A_{|110\rangle}-\Gamma_1A_{|010\rangle},\\
    \frac{dA_{|100\rangle}}{dt}&=\Gamma_{10} A_{|110\rangle}-\Gamma_{10} A_{|100\rangle},\\
    \frac{dA_{|000\rangle}}{dt}&=\Gamma_{10} A_{|010\rangle}+\Gamma_{10} A_{|100\rangle},
\end{align}
with initial condition $A_{|120\rangle}(t=0)=1$,
yielding:
\begin{align}
    A_{|120\rangle}&=e^{-(\Gamma_{10}+\Gamma_{21})t},\\
    A'_{|120\rangle} &= \frac{A_{|120\rangle}}{A_{|120\rangle}+A_{|020\rangle}+A_{|100\rangle}+A_{|000\rangle}}=\frac{1}{e^{(\Gamma_{10}+\Gamma_{21})t}+\lambda},\label{a120} \\
    \lambda &= \frac{\Gamma_{21}(e^{\Gamma_{10} t}-e^{\Gamma_{21} t})}{\Gamma_{21}-\Gamma_{10}},
\end{align}
where $A'_{|120\rangle}$ represents the probability amplitude of the state $|120\rangle$ after PS, 
and $\lambda$ in the denominator accounts for the effect of PS. 
Similarly, the target state evolution in the system with the switch $\{|0\rangle,|1\rangle\}$ is given by:
\begin{gather}
    A'_{|110\rangle}=e^{-2\Gamma_{10} t}.\label{a110}
\end{gather}

\begin{figure}[ht]
    \centering
    \includegraphics[width=0.6\linewidth]{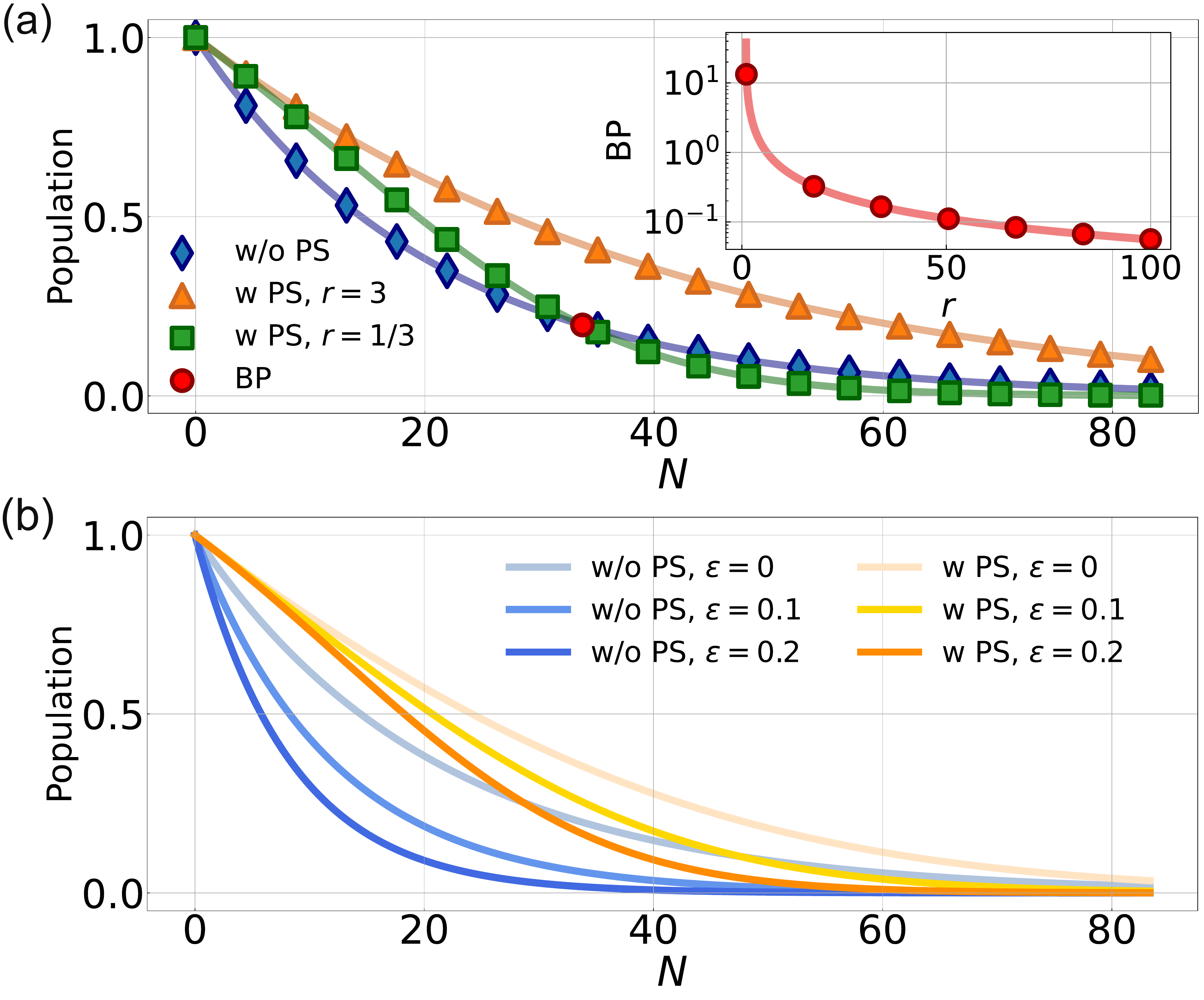}
    \caption{\justifying
        (a) Comparison of schemes under different ratios of \(\Gamma_{10}\) and \(\Gamma_{21}\) (\(r = \Gamma_{21}/\Gamma_{10}\)). 
        We set \(\Gamma_{10} = 1/15~\mu\mathrm{s}^{-1}\) and a CSWAP duration of \(90\)~ns. 
        Points represent simulation data obtained using QuTiP for \( {\rm CSWAP}^{2N}\)~\citesm{QUTIP}, while lines represent theoretical predictions. 
        The BP (marked in red) reflects the equilibrium position with and without the PS scheme when \(r > 1\). 
        The inset shows the relationship between BP and \(r\). 
        As \(r\) increases, the BP position progressively decreases until it reaches the point where \(N = 0\).
        (b) Comparison of different schemes considering the leakage rate (\(\epsilon\)). Here, we focus on the case where \(r = 1.2 \) and examine scenarios with \(\epsilon = 0\), \(0.1\), and \(0.2\). 
        As \(\epsilon\) increases, the decay becomes progressively more severe in both schemes.
        However, the PS scheme retains its advantage over the original scheme and demonstrates greater resilience to leakage, as evidenced by the comparison of population at the same position where $N<40$.  
    }
    \label{fig deconhSim}
\end{figure}

By taking the difference between equations Eq.~\eqref{a110} and Eq.~\eqref{a120}, we obtain three zero points:
\begin{align}
    t_1 &= 0,\\
    t_2 &= \infty,\\
    t_{3} &=-\frac{\log(1-\Gamma_{10}/\Gamma_{21})}{\Gamma_{10}},\ when\  \Gamma_{10}<\Gamma_{21},
\end{align}
where \(t_1\) and \(t_2\) reflect the consistency between the initial and final states of the system in the two schemes.
The third zero point, \(t_3\), appears if and only if \(\Gamma_{10} < \Gamma_{21}\),
representing the balance point (BP) where the eraser technique outperforms the non-eraser scheme, as illustrated in Fig.~\ref{fig deconhSim}(a).  
The BP serves as a crucial metric for determining appropriate practical setups for the system.  
If \(\Gamma_{10} > \Gamma_{21}\), the PS technique consistently outperforms the original scheme.  
Both schemes exhibit equivalent decoherence behavior due to the equivalence of second-order and first-order processes.
We compared simulation data for different decoherence parameters in Fig.~\ref{fig deconhSim}(b).

We also consider the leakage introduced by the imperfect \(\sqrt{\rm CZ}\) gate, modeled by rate $\epsilon$.  
In the non-eraser scheme, the leakage manifests as transitions to the state $|020\rangle$, 
while it manifests as transitions to the state $|111\rangle$ in the eraser scheme.  
The latter can be detected and filtered out through PS, whereas the former persists.  
The corrected population expressions are:
\begin{align}
    A''_{|110\rangle}&=\epsilon^{-(2\Gamma_{10}+\epsilon)t},\\
    A''_{|120\rangle}&=\frac{(\Gamma_{10} - \Gamma_{21})(\Gamma_{10} + 	\epsilon)(\Gamma_{21} + \epsilon)(\Gamma_{10} + \Gamma_{21} + \epsilon)}{M_1+M_2+M_3+M_4},
\end{align}
with:
\begin{align}
    M_1 &= (\Gamma_{10}^3 \Gamma_{21} - \Gamma_{10} \Gamma_{21}^3 + \Gamma_{10}^3 \epsilon + \Gamma_{10}^2 \Gamma_{21} \epsilon - \Gamma_{10} \Gamma_{21}^2 \epsilon + \Gamma_{10}^2 \epsilon^2 - \Gamma_{21}^3 \epsilon - \Gamma_{21}^2 \epsilon^2)e^{(\Gamma_{10}+\Gamma_{21}+\epsilon)t},\\
    M_2&= (-\Gamma_{10}^2 \Gamma_{21}^2 - \Gamma_{10} \Gamma_{21}^3 - \Gamma_{10}^2 \Gamma_{21} \epsilon - 2 \Gamma_{10} \Gamma_{21}^2 \epsilon - \Gamma_{10} \Gamma_{21} \epsilon^2)e^{(\Gamma_{10}+\epsilon)t},\\
    M_3 &= (\Gamma_{10}^2 \Gamma_{21}^2 + \Gamma_{10} \Gamma_{21}^3 + 2 \Gamma_{10} \Gamma_{21}^2 \epsilon + \Gamma_{21}^3 \epsilon + \Gamma_{21}^2 \epsilon^2) e^{(\Gamma_{21}+\epsilon)t},\\
    M_4 &= 2 \Gamma_{10}^2 \Gamma_{21} \epsilon - \Gamma_{10} \Gamma_{21}^2 \epsilon - \Gamma_{21}^3 e + \Gamma_{10}^2 \epsilon^2 + \Gamma_{10} \Gamma_{21} \epsilon^2 - 2 \Gamma_{21}^2 \epsilon^2 + \Gamma_{10} \epsilon^3 - \Gamma_{21} \epsilon^3.
\end{align}

Simulation results in Fig.~\ref{fig deconhSim}(b) confirm the resilience of the PS scheme to leakage.
Although coherent errors may obscure the BP, it remains a valuable metric in idealized analysis.
Ultimately, the growing impact of noise at increasing circuit depths necessitates more advanced error-mitigation techniques for a scalable routing system.
It is also crucial to develop robust strategies to mitigate the effects of decoherence and noise in practical QRAM systems.

\subsection{Simulation results about leakage in a QRouter}{\label{leakage}}

In practical experimental implementations, the $\rm \sqrt{CZ}$ encounters challenges in achieving perfect population transfer between the states \( |11\rangle \) and \( |20\rangle \) due to technical issues such as distortion and crosstalk. Consequently, the implemented unitary in Eq.~\eqref{CP} effectively becomes an imperfect $\mathrm{iSWAP}_{02\leftrightarrow 11}(\vartheta \neq \pi, \eta)$.
For simplicity, we first consider $\sqrt{\mathrm{CZ}'} = \mathrm{iSWAP}_{02\leftrightarrow 11}(\vartheta, \eta=0)$. The resulting CSWAP$'$ gate, acting in the Hilbert subspace spanned by \(\{|011\rangle, |020\rangle, |110\rangle, |111\rangle, |021\rangle, |120\rangle\}\), is given by:
 \begin{equation}\label{eq:cswap_leakage}
     \begin{aligned}
        {\rm CSWAP'} &=
        \begin{pmatrix}
            \cos{(\vartheta/2)} & -i\sin{(\vartheta)}/2 & -\sin{(\vartheta/2)}^2 & 0 & 0 & 0 \\
            -i\sin{(\vartheta)}/2 & K_1 & K_2 & 0 & 0 & 0 \\
            -\sin{(\vartheta/2)}^2 & K_2 & K_3 & 0 & 0 & 0 \\
            0 & 0 & 0 & K_1 & K_2 & -i\sin{(\vartheta)}/2 \\
            0 & 0 & 0 & K_2 & K_3 & -i\sin{(\vartheta)}/2 \\
            0 & 0 & 0 & -i\sin{(\vartheta)}/2 & -\sin{(\vartheta/2)}^2 & \cos{(\vartheta/2)} \\
        \end{pmatrix},
     \end{aligned}
\end{equation}
where
\begin{equation}
     \begin{aligned}
        K_1 &= \cos{(\vartheta/2)}^3-\sin{(\vartheta/2)}^2,\\
        K_2 &= -i\sin{(\vartheta)}-i\cos{(\vartheta/2)}\sin{(\vartheta)},\\
        K_3 &= cos{(\vartheta/2)}^2-\cos{(\vartheta/2)}\sin{(\vartheta)}^2.
     \end{aligned}
\end{equation}
We also observe that the phase of the address qutrit's state \( |2\rangle \) influences leakage accumulation in the subspace \(\{|011\rangle, |020\rangle, |110\rangle\}\).
In the space $\{|111\rangle, |021\rangle, |120\rangle\}$,
the influence of address qutrit's state \( |1\rangle \) is similar.
These phase shifts often originate from AC Stark shifts induced by microwave drives or direct qubit frequency modulation. Taking \(X_{\pi}^{01}\) and \( X_{\pi}^{12}\) as examples, their mathematical descriptions in \(\{|0\rangle, |1\rangle, |2\rangle\) are given by:
\begin{equation}\label{eq:X_2phase}
     \begin{aligned}
        {X_{\pi}^{01}(\phi')} &=
        \begin{pmatrix}
            0 & 1 & 0 \\
            1 & 0 & 0 \\
            0 & 0 & e^{i\phi'} \\
        \end{pmatrix},\\
        {X_{\pi}^{12}(\phi'')} &=
        \begin{pmatrix}
            e^{i\phi''} & 0 & 0 \\
            0 & 0 & 1 \\
            0 & 1 & 0 \\
        \end{pmatrix},
     \end{aligned}
\end{equation}
where $\phi'$ and $\phi''$ are the additional phases on states \( |2\rangle \) and \(|1\rangle \) introduced by the non-resonant microwave actuation, respectively.

\begin{figure}[ht]
    \centering
    \includegraphics[width=0.6\linewidth]{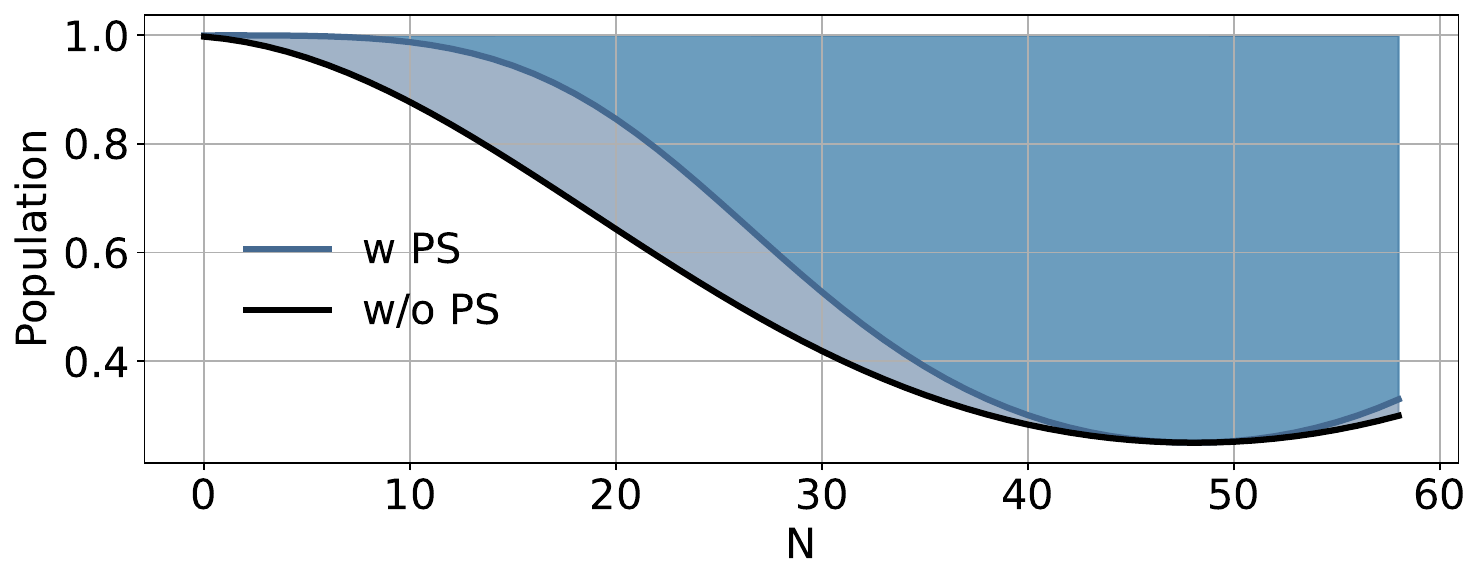}
    \caption{\justifying
    Simulation results about leakage when $\vartheta=0.99\pi$. The solid line represents the most severe leakage case, where the phase (\(\phi'\), \(\phi''\)) is $\pi/2$,
    corresponding to the constructive interference situation.
    When the phase is 0, the interference is in a destructive state, and the leakage effect is minimal.
    }
    \label{fig leakage}
\end{figure}

Subsequently, 
we consider the QRouter configuration in Eq.~\eqref{eq:QR} and perform a leakage simulation analysis. The initial state is set to \( |1100\rangle \) (\( |1200\rangle \)), and we observe the population of this state under the non-eraser scheme (eraser scheme) after two applications of the QRouter.
Simulation results are presented in Fig.~\ref{fig leakage}.

Figure~\ref{fig leakage} respectively illustrates the cases under the eraser scheme (blue) and the non-eraser scheme (black). The shaded region reflects variation in \(\phi'\) and \(\phi''\) from $0$ to \(\pi/2\).
Leakage accumulation exhibits phase-dependent interference effects, which modulate the information routing behavior of the QRouter.
Specifically, leakage affects the transmission ratio of the target path and can erroneously redirect information to the wrong path. When the phase is set to \(\pi/2\), constructive interference maximizes leakage, manifesting as periodic coherent errors. In contrast, setting the phase to 0 suppresses leakage significantly. Post-selection can further enhance fidelity by excluding erroneous outcomes. Thus, leakage can be actively mitigated through phase control, either via physical \(Z\) gates or virtual phase gates, as well as through post-processing techniques.

\subsection{Detailed numerical simulations}{\label{plus}}

To assess the combined effects of decoherence and leakage on QRouter performance, we conduct full numerical simulations of the Random Access Test (RAT), incorporating both coherent leakage and incoherent noise. Decoherence is modeled using a qutrit-specific formalism from Ref.~\citesm{morvan2021qutrit}, which is given by:

\begin{align}
\begin{pmatrix}
\rho_{00}(t) \\
\rho_{01}(t) \\
\rho_{02}(t) \\
\rho_{10}(t) \\
\rho_{11}(t) \\
\rho_{12}(t) \\
\rho_{20}(t) \\
\rho_{21}(t) \\
\rho_{22}(t)
\end{pmatrix}
=
\begin{pmatrix}
1 & 0 & 0 & 0 & 1 - e^{-\Gamma_{10} t} & 0 & 0 & 0 & V_1 \\
0 & e^{-\Gamma_{2} t} & 0 & 0 & 0 & 0 & 0 & 0 & 0 \\
0 & 0 & e^{-\Gamma_3 t} & 0 & 0 & 0 & 0 & 0 & 0 \\
0 & 0 & 0 & e^{-\Gamma_{2} t} & 0 & 0 & 0 & 0 & 0 \\
0 & 0 & 0 & 0 &  e^{-\Gamma_{10} t} & 0 & 0 & 0 & V_2 \\
0 & 0 & 0 & 0 & 0 & e^{-\Gamma_{4} t} & 0 & 0 & 0 \\
0 & 0 & 0 & 0 & 0 & 0 & e^{-\Gamma_3 t} & 0 & 0 \\
0 & 0 & 0 & 0 & 0 & 0 & 0 & e^{-\Gamma_4 t} & 0 \\
0 & 0 & 0 & 0 & 0 & 0 & 0 & 0 & e^{-\Gamma_{21} t}
\end{pmatrix}
\begin{pmatrix}
\rho_{00}(t) \\
\rho_{01}(t) \\
\rho_{02}(t) \\
\rho_{10}(t) \\
\rho_{11}(t) \\
\rho_{12}(t) \\
\rho_{20}(t) \\
\rho_{21}(t) \\
\rho_{22}(t)
\end{pmatrix}, 
\end{align}
where
\begin{align}
V_1&=1-\frac{\Gamma_{10}e^{-\Gamma_{21} t} - \Gamma_{21}e^{-\Gamma_{10} t}}{\Gamma_{10} - \Gamma_{21}},\\
V_2&=\frac{\Gamma_{21} \left(e^{-\Gamma_{21} t} - e^{-\Gamma_{10} t}\right)}{\Gamma_{10} - \Gamma_{21}}.
\end{align}
We denote the rates of the decay processes by $\Gamma_{21} = 1/T^{(21)}_{1}$ and $\Gamma_{10} = 1/T^{(10)}_{1}$,
while the the rates of the echo dephasing processes are $\Gamma_{2} = 1/T^{(10)}_{2}$, $\Gamma_{3} = 1/T^{(21)}_{2}$, and $\Gamma_{4} = 1/T^{(20)}_{2}$.
Here, $T_1$ and $T_2$ parameters used to describe qubit decoherence.
In our processor, the typical values at the qubit working frequencies are  $\Gamma_{10} = 1/15~{\mu\mathrm{s}}^{-1}$,  $\Gamma_{21} = 1/12~{\mu\mathrm{s}}^{-1}$, $\Gamma_{2} = \Gamma_{3}=\Gamma_{4}=1/2.5~{\mu\mathrm{s}}^{-1}$.

\begin{figure}[ht]
\centering	        
\includegraphics[width=0.6\linewidth]{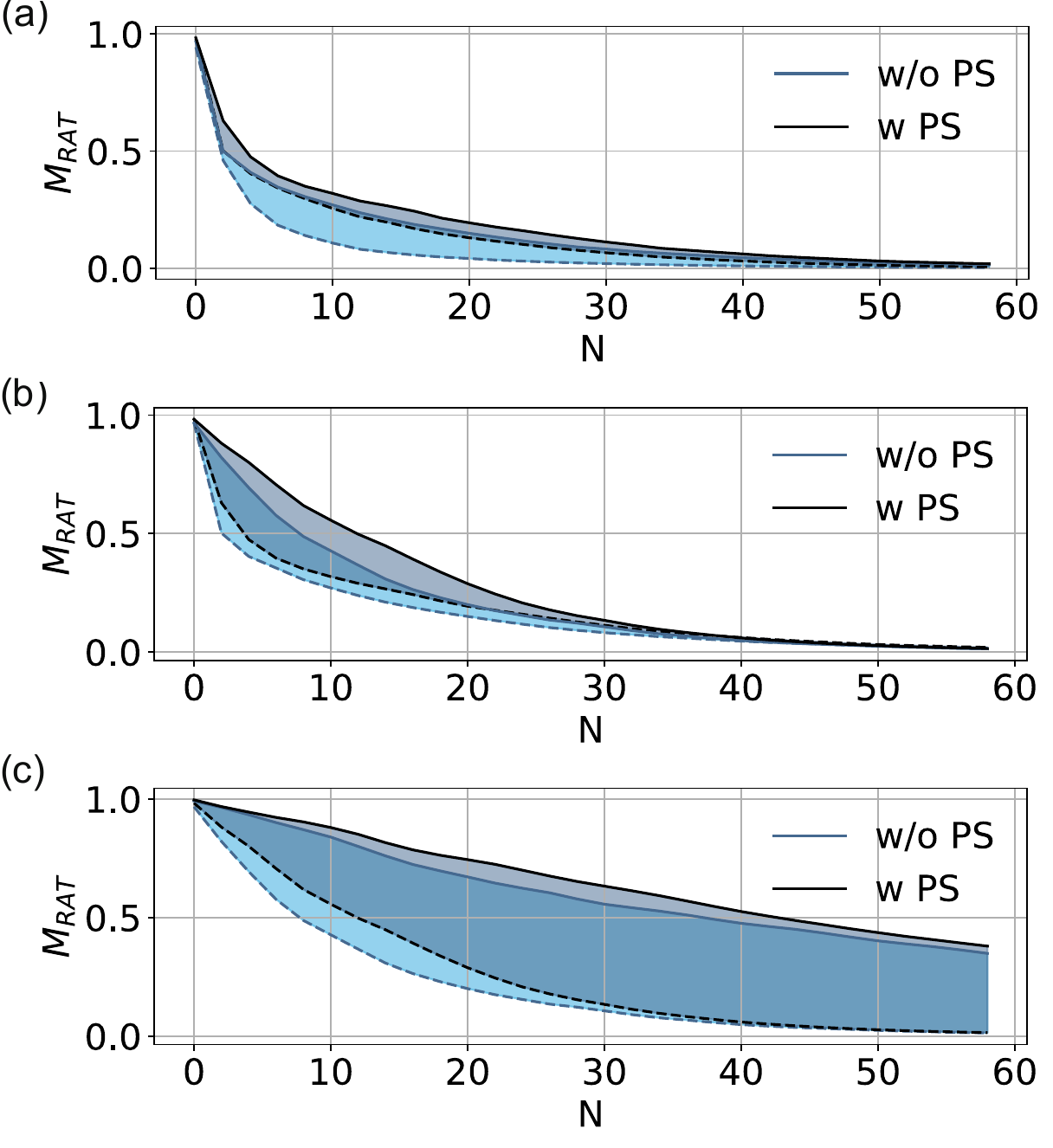}
\caption{\justifying
Simulation results of RAT under varying (a) leakage error, (b) \( T_2 \), and (c) \( T_1 \) .
The shaded areas correspond to parameter sweeps. Dashed lines represent minimal fidelity, and solid lines correspond to optimal settings.
    }
    \label{fig sim_de}
\end{figure}

We actively configured system parameters and conducted random access test (RAT) simulations involving 100 groups of random accesses with a circuit depth of 60, to individually evaluate the impact of leakage, \( T_2 \), and \( T_1 \) on system fidelity. 
First, leakage was found to have a pronounced effect. Under typical system decoherence, we varied the leakage rate and analyzed the resulting RAT fidelity, as shown in Fig.~\ref{fig sim_de}(a). When the leakage rate ($\delta \vartheta$) was reduced from 5\% (dashed line) to 0\% (solid line), the RAT fidelity \( F^{\rm w/o~PS}_{\rm RAT} \) (non-eraser) and \( F^{\rm w~PS}_{\rm RAT} \) (eraser) improved from 86.18\% and 93.07\%  to 93.42\% and 94.81\%, respectively. The dark blue shaded region corresponds to results obtained under eraser scheme, while the light blue area reflects results under non-eraser scheme.
Dephasing, governed by \( T_2 \), is another key contributor to fidelity degradation. Figure~\ref{fig sim_de}(b) illustrates the effect of increasing \( T_2 \) from 2.5~$\mu$s (dashed line) to $2T_1$ (solid line), during which the fidelity curves exhibit a shift in concavity—from steep to flat. This indicates a  reduced rate of fidelity loss and results in fidelity improvements from 93.42\% (non-eraser) and 94.81\% (eraser) to 95.99\% and 97.15\%, respectively. Relaxation, characterized by \( T_1 \), also significantly influences performance. We held \( T_2 = 2T_1 \) and maintained a fixed ratio \(T^{(10)} = 1.25T^{(21)} \), then increased \( T_1 \) from 15~$\mu$s (dashed line) to 60~$\mu$s (solid line). As shown in Fig.~\ref{fig sim_de}(c), higher \( T_1 \) values reduce the curve's decay rate, leading to enhanced fidelity—from 95.99\% (non-eraser) and 97.14\% (eraser) to 98.82\% and 99.51\%, respectively.
In summary, minimizing leakage and ensuring sufficiently long decoherence times enable our QRouter to achieve high fidelity, meeting the demands of large-scale QRAM construction and deployment.

\section{Unidirectional quantum routing network based on the short path design}{\label{sec:levels11}}

\begin{figure}[ht]
\centering
\includegraphics[width=0.5\linewidth]{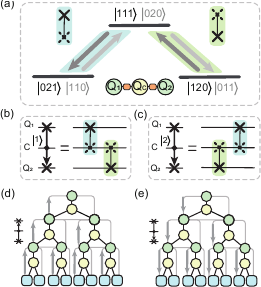}
\caption{\justifying
    (a) Transition pathways in the short-path TCG (SP-TCG) scheme for the short-path CSWAP (SP-CSWAP) operation. Unlike the full CSWAP operation, the evolution in SP-CSWAP operation is unidirectional, e.g., $|110\rangle \rightarrow |020\rangle \rightarrow |011\rangle$, and $|021\rangle \rightarrow |111\rangle \rightarrow |120\rangle$.
    (b) Circuit implementation of SP-CSWAP with control basis ${|0\rangle, |1\rangle}$.
    (c) Circuit implementation of SP-CSWAP with control basis ${|0\rangle, |2\rangle}$.
    (d)-(e) Schematic diagrams of QWOM and QROM, respectively.
    The QRouter is realized through the operations shown in (b) and (c), enabling unidirectional data transfer in QWOM and QROM.
    }
    \label{fig SPTCG}
\end{figure}

This section presents the design of unidirectional quantum routers (QRouters) implemented via the short-path transition controlled gate (SPTCG) scheme~\citesm{zhang2024realization}. 
The SPTCG scheme employs a non-unitary evolution process to achieve unidirectional state transitions from one initial state to the target state. Compared to the full TCG scheme, SPTCG scheme utilizes only approximately half of the available transition pathways, thereby reducing the required quantum resources and simplifying circuit implementation. Since the address loading and unloading process in QRAM inherently involves unidirectional information transfer, constructing QRouters based on SPTCG leads to circuits with significantly reduced depth.
Following this idea, we introduce the short-path controlled SWAP (SP-CSWAP) operation, as illustrated in Fig.~\ref{fig SPTCG}. This operation is characterized by asymmetric structure and transition pathways, which facilitate unidirectional data flow. The use of the qutrit states $|0\rangle$ and $|2\rangle$ as address levels, together with eraser techniques, continues to enhance gate fidelity. Notably, the SP-CSWAP operations using address bases ${|0\rangle, |1\rangle}$ and ${|0\rangle, |2\rangle}$ are inverses of each other due to their opposite transition directions.

Furthermore, the SP-CSWAP operations enable two specialized variants of QRAM: quantum read-only memory (QROM) and quantum write-only memory (QWOM)~\citesm{paler2020parallelizing}. QROM facilitates one-way data transfer from the database to the data bus, while QWOM performs the reverse operation. For these targeted use cases, SP-TCG-based circuits can be further optimized to minimize circuit depth. Compared to complete unitary implementations, the short-path approach reduces circuit depth by approximately 66\%, thereby offering improved scalability. This architecture holds promising potential for large-scale quantum information processing and specialized quantum computing applications.


\let\oldbibitem\bibitem
\renewcommand{\bibitem}[2][]{
  \ifstrempty{#1}{\oldbibitem{SM_#2}}{\oldbibitem[#1]{SM_#2}}
}

\end{document}